\documentclass[12pt,a4paper, else]{article}
\usepackage{PRIMEarxiv}

\usepackage[utf8]{inputenc}
\usepackage{alphabeta} 
\usepackage{geometry} 

\usepackage[parfill]{parskip}

\newcommand{\eat}[1]{}

\usepackage{enumitem}
\usepackage{bbm}
\setlist{nolistsep,noitemsep}
\usepackage[hidelinks]{hyperref}

\usepackage[greek,english]{babel}
\usepackage{cite}
\usepackage{lipsum}
\usepackage{lastpage}
\usepackage{fancyhdr}
\usepackage{algorithmic}
\usepackage{subcaption}
\usepackage{placeins}
\usepackage{booktabs}
\usepackage{xcolor}
\usepackage{mathtools}
\usepackage{wrapfig}
\usepackage{amssymb}
\usepackage{bm}
\hypersetup{
  colorlinks   = true, 
  urlcolor     = blue, 
  linkcolor    = blue, 
  citecolor   = blue 
}

\usepackage{amsmath,amsfonts}
\usepackage[noend,ruled,vlined]{algorithm2e}
\usepackage{array}

\usepackage{lipsum}
\usepackage[nameinlink]{cleveref}

\RequirePackage[normalem]{ulem} 
\RequirePackage{color}\definecolor{RED}{rgb}{1,0,0}\definecolor{BLUE}{rgb}{0,0,1} 
\RequirePackage{listings} 
\RequirePackage{color} 
\lstdefinelanguage{DIFcode}{ 
  moredelim=[il][\color{red}\sout]{\%DIF\ <\ }, 
  moredelim=[il][\color{blue}\uwave]{\%DIF\ >\ } 
} 
\lstdefinestyle{DIFverbatimstyle}{ 
	language=DIFcode, 
	basicstyle=\ttfamily, 
	columns=fullflexible, 
	keepspaces=true 
} 
\lstnewenvironment{DIFverbatim}{\lstset{style=DIFverbatimstyle}}{} 
\lstnewenvironment{DIFverbatim*}{\lstset{style=DIFverbatimstyle,showspaces=true}}{} 

\usepackage[some]{background}
    \backgroundsetup{%
      scale=1,       
      angle=0,       
      opacity=0.7,    
      color = {darkgray},  
      contents={\parbox{\textwidth}{\centering 
     \large This paper was accepted for publication in Applied Energy \\[25cm]}
    }
}

\title{Vehicle-to-grid for car sharing - A simulation study for 2030}

\author{
  Nina Wiedemann \\
  Institute of Cartography and Geoinformation \\
  ETH Zurich, Switzerland \\
  \texttt{nwiedemann@ethz.ch} \\
   \And
  Yanan Xin \\
  Institute of Cartography and Geoinformation \\
  ETH Zurich, Switzerland \\
  \AND
  Vasco Medici \\
  SUPSI, Mendrisio, Switzerland \\
  \And
  Lorenzo Nespoli \\
  SUPSI, Mendrisio, Switzerland \\
  Hive Power SA, Suglio, Switzerland \\
  \And 
  Esra Suel \\
  The Bartlett Centre for Advanced Spatial Analysis \\
  University College London, UK
  \And
  Martin Raubal \\
  Institute of Cartography and Geoinformation \\
  ETH Zurich, Switzerland \\
}

\begin{document}


\maketitle

\begin{abstract}
The proliferation of car sharing services in recent years presents a promising avenue for advancing sustainable transportation. Beyond merely reducing car ownership rates, these systems can play a pivotal role in bolstering grid stability through the provision of ancillary services via vehicle-to-grid (V2G) technologies - a facet that has received limited attention in previous research. In this study, we analyze the potential of V2G in car sharing by designing future scenarios for a national-scale service in Switzerland.
We propose an agent-based simulation pipeline that considers population changes as well as different business strategies of the car sharing service, and we demonstrate its successful application for simulating scenarios for 2030. To imitate car sharing user behavior, we develop a data-driven mode choice model.
Our analysis reveals important differences in the examined scenarios, such as higher vehicle utilization rates for a reduced fleet size as well as in a scenario featuring new car sharing stations. These disparities translate into variations in the power flexibility of the fleet available for ancillary services, ranging from 12 to 50 MW, depending on the scenario and the time of the day. Furthermore, we conduct a case study involving a subset of the car sharing fleet, incorporating real-world electricity pricing data. The case study substantiates the existence of a sweet spot involving monetary gains for both power grid operators and fleet owners. 
Our findings provide guidelines to decision makers and underscore the pressing need for regulatory enhancements concerning power trading within the realm of car sharing. 
\end{abstract}


\BgThispage{}

\section{Introduction}\label{sec:intro}


The enormous challenges the world is facing due to climate change necessitate the transformation of entire industries, especially the energy and transportation sectors. Innovative solutions are imperative for a substantial reduction in CO\textsubscript{2} emissions. Within the energy sector, renewable energies are increasingly adopted, but introduce new challenges, such as strong fluctuations at limited storage capacity. In the transportation sector, electric vehicles (EVs) are the preferred alternative but involve high electricity demand. Furthermore, replacing all personal vehicles with EVs will arguably not suffice, due to the high CO\textsubscript{2} emissions in vehicle production. New mobility concepts such as Mobility as a Service (MaaS), micromobility~\cite{reck_explaining_2021}, and shared electric vehicles, present viable alternatives that reduce production and usage emissions while continuing to provide individual transport options. 

Shared EV fleets can, in addition, support the transition to renewable energies by offering ancillary services through smart charging or vehicle-to-grid (V2G) technologies. The concept of V2G has received much attention in recent years~\cite{raviUtilizationElectricVehicles2022}, as it simultaneously supports sustainable transportation and robust renewable energy supply. \cite{taiebat2019synergies} explore emerging technologies in this domain and explicitely highlight the combination of V2G with shared (automated) vehicles as a promising opportunity that requires further quantitative analysis. V2G integration within car sharing presents several advantages: 1) The car sharing service owner can coordinate and manage vehicle availability to effectively integrate V2G technology, in contrast to less predictable driving choices of private vehicle owners, 2) the large fleet size in car sharing services, sometimes comprising more than thousand vehicles, enables sizable profits from implementing V2G and substantial benefits to the grid, and 3) the rental pricing structure can incentivize vehicle usage for mobility and charging outside peak demand. Notably, a stated-choice experiment by \cite{gschwendtner2022coupling} indicates that car sharing users are generally interested in supporting V2G initiatives. Nevertheless, previous research on V2G has predominantly focused on privately owned cars or commercial fleets~\cite{tepe2022optimal} rather than car sharing. For example, they analyze the potential of V2G of private vehicles under varying energy production and consumption patterns~\cite{dallinger2011vehicle, loisel2014large, ji2020evaluating, knupfer2016cross, sassi2021vehicle, cai2022optimizing, lauvergne2022integration}. The car sharing literature, on the other hand, mainly evaluates the impact of car sharing on car ownership rates~\cite{mishra2015effect, martin2011impact, liao2020carsharing}, resource efficiency~\cite{glotz2016reclaim}, CO\textsubscript{2} emissions~\cite{sun2018completive}, and transport mode choices~\cite{ayed2015using, balac2015carsharing, ciari2013estimation}. 
Only a limited number of studies have integrated V2G or smart charging with car sharing, primarily within the context of optimizing the charging and relocation of car sharing fleets~\cite{zhang_values_2020, caggiani_static_2021, xu_electric_2021, nespoli2022national}. These algorithms are usually tested on small fleets instead of realistic car sharing systems; only \cite{nespoli2022national} have recently optimized V2G for a national-scale car sharing fleet. Meanwhile, there remains a lack of studies evaluating \textit{future} opportunities of V2G in car sharing. Given the rapid adoption of electric vehicles and the potential expansion of car sharing services in the near future, it is critical to estimate the potential monetary benefits and peak-shaving potential of V2G in large car sharing fleets with respect to possible future scenarios. These estimations are essential to guide policymaking and accelerate the electrification process of the mobility sector.

We fill this gap with a simulation study of V2G operations in future car sharing scenarios. 
Given the prospect of high car sharing prevalence in 2030 and beyond~\cite{zhou2017projected}, as well as its potential growth with the rise of autonomous driving~\cite{narayanan2020shared}, it is paramount to develop a better understanding of its compatibility with V2G. However, making long-term projections of car sharing is challenging since there are a multitude of influencing factors, including the growth of car sharing services and their customer base, shifts in mobility behavior, the general adoption of EVs, and the pricing dynamics of ancillary services. 
Here, we propose and implement an agent-based simulation to capture the intricacies of mobility behavior. Our approach to modeling transport mode choices is data-driven and relies on machine learning techniques. The simulation approach involves two key components: firstly, the scaling of the car sharing system to simulate the additional number of vehicles and stations in the future, and, secondly, the expansion of the customer base through the creation of a synthetic population derived from census data. Finally, we apply a V2G optimization algorithm to the simulated car sharing data to quantify the potential for peak-shaving and the associated monetary savings for the car sharing service. 

To simulate a realistic and large-scale car sharing service, we leverage a dataset of one of the largest currently operating car sharing fleets, run by the Swiss company \textit{Mobility}. We find that additional car sharing stations induce more car sharing demand than simply adding more vehicles at existing car sharing stations. Furthermore, a larger fleet leads to higher charging and discharging flexibilities. 
We further conducted a case study with real electricity pricing data from a regional power grid operator over all simulated future scenarios. The case study demonstrates that V2G in car sharing can provide a substantial peak-shaving effect given a reasonable price for ancillary service provision, resulting in a win-win situation for both fleet and grid operators.

Our contributions are summarized in the following:
\begin{itemize}
    \item Presenting a full pipeline to simulate future car sharing booking patterns at the national population level 
    \item Learning car sharing user mode choice behavior with a machine learning-based mode choice model
    \item Developing algorithms to simulate station layout and booking behavior
    \item Analyzing future opportunities for V2G technology to provide ancillary services in car sharing businesses
\end{itemize}

In the following, we will discuss related work on car sharing simulation and V2G scenarios in \autoref{sec:background}, before presenting our methods in \autoref{sec:methods}. \Cref{sec:res} presents our experiments, where our car sharing simulation is first validated, then applied to analyze car sharing behaviour in six scenarios for 2030, and finally combined with V2G optimization. We conclude with discussion and outlook in \autoref{sec:discussion} and \autoref{sec:conclusion}.

\section{Related work}\label{sec:background}

\subsection{Car sharing services}

Car sharing services have evolved rapidly in recent years~\cite{shaheen_carsharing_2013}, creating numerous novel research opportunities~\cite{ferrero_car-sharing_2018}. To simulate future car sharing behavior, it is necessary to assess the characteristics of (potential) target groups. For Switzerland, \cite{juschten_carsharing_2019}, for instance, discuss factors influencing car sharing membership rates, such as proximity to public transport. \cite{amirnazmiafshar_review_2022} and \cite{becker_comparing_2017} concentrate on sociodemographic variables and find that the proximity of the living location to car sharing stations is a critical factor, alongside age, gender, mobility behavior, car sharing membership and usage patterns. These insights are implicitly used in predictive models (e.g. \cite{cocca_car-sharing_2020}) that aim to assist car sharing operators by estimating future demand or potential demand at new stations~\cite{kumar2012optimizing, muhlematter2023spatially}. Being a niche market accounting for less than 0.1\% of transportation activities in Switzerland, car sharing is hardly considered in transport mode choice models, with few exceptions~\cite{kim2017effects}. 

%
%
%

\subsection{Car sharing simulation}

To model car sharing behavior in 2030, we consider car sharing simulators proposed in previous work. 
Agent-based models have been extensively used in transportation as they provide a realistic representation of human society, but they are oftentimes computationally expensive.
%
%
%
For car sharing specifically, the powerful MATSim traffic simulator has been extended to both return- or one-way car sharing trips~\cite{ciari_modeling_2016, ayed2015using, balac2015carsharing, ciari2013estimation}. For example, \cite{giorgione2020dynamic} use MATSim to evaluate dynamic pricing schemes for car sharing in Berlin. However, MATSim is primarily suited for capturing competition between transport modes via their respective utilities, and to model the effects of interventions on a macroscopic level. In this study, we aim to model realistic car sharing mode choice behavior with respect to the characteristics of a future population and infrastructure, which is not captured sufficiently in the utility function that determines agents' choices in MATSim. Additionally, running a MATSim simulation for the whole Swiss population is disproportionately computationally expensive, considering the low share of car sharing among other transport modes. 

In contrast, event-based simulators are more efficient since they avoid modeling individual agents in the system. Instead, they model the spatial and temporal distribution of car sharing bookings via sampling. \cite{cocca_free_2019, cocca_car-sharing_2020} and \cite{fassio_environmental_2021} propose an event-based car sharing simulation approach in which they utilize a Poisson process to model the temporal distribution and Kernel Density Estimation for the spatial distribution of car sharing pick-ups or drop-offs. 
However, the transferability of such a model to simulating future scenarios remains unclear. Relying solely on the statistical distribution of events within the current car sharing service, without considering population change, simulating future scenarios would essentially entail making arbitrary alterations to event rates and other statistical parameters. Since an agent-based approach is essential to account for the evolving sociodemographics of the population, we develop an agent-based simulator with improved efficiency over MATSim by omitting transport modes other than car sharing. In addition, our simulator includes a data-driven mode choice model to simulate car sharing usage behaviors.


\subsection{Future scenarios}

Scenario-based analysis is an important research tool to explore potential outcomes or determine pathways toward a desirable future, offering a long-term view under uncertainties~\cite{schoemaker1995scenario, schwartz2012art, abbott2005understanding}.
\cite{abou2022overview} summarize scenario-based research and, building upon the study by \cite{borjeson2006scenario}, distinguish between predictive/probable scenarios, normative/preferable scenarios, and exploratory scenario planning (XSP). Here, we focus on XSP, aiming to explore alternative future developments from various viewpoints~\cite{abou2022overview}. XSP requires identifying driving forces for future development and important stakeholders, selecting factors of uncertainty, and discretizing their values in a scenario matrix~\cite{stapletonexploratory}. Most commonly, 3-5 scenarios are constructed to ensure sufficient variety while limiting the complexity of the analysis
~\cite{avin2022exploratory}. 

Furthermore, scenario planning was integrated with road mapping for business applications~\cite{strauss2004roadmapping, gerdsri2007analytical, saritas2010using}. \cite{geum2014combining} extend these theoretical considerations with a practical framework for combining a simulation of the system dynamics with scenarios from the technological and business perspectives. We build upon their work since \cite{geum2014combining} demonstrate the applicability of their framework for constructing car sharing service scenarios.

\subsection{Analyses of the future potential of V2G}

Among studies that evaluate future perspectives for V2G, the main differences lie in the complexity of the simulation of mobility behavior and the assumptions on charging opportunities. Most studies project the energy consumption and production patterns of a population, and simulate V2G operations to evaluate the potential impact on the grid, e.g. for Germany~\cite{loisel2014large, hartmann2011impact}, Morocco~\cite{sassi2021vehicle}, the UK~\cite{knupfer2016cross} or China~\cite{ji2020evaluating}. \cite{sassi2021vehicle}, for instance, find that V2G can provide up to 7.7 GW of controllable and mobile loads for the Moroccan grid in 2030. \cite{wang2021value} argue that there are ``four factors that could drive future V2G revenues: future grid changes, large EV numbers, V2G interactions with electricity prices, and V2G operational responses to shifts in electricity prices" (p. 1).  They claim to consider all four factors in their analysis of future revenues from V2G in California by 2030. However, such studies may still be misleading as they oftentimes disregard the potential change in people's mobility behavior. \cite{dallinger2011vehicle} distinguish \textit{static} and \textit{dynamic} simulations, where the energy demand is given as daily averages or as dynamic mobility patterns respectively. They show that the dynamic scenario reduces the power available for ancillary services by 40\%. \cite{martin2022using} tackle this issue by using tracking data in their case study on smart charging with photovoltaic-generated energy, and \cite{xu2018planning} combine data from charging stations and mobile phone data to account for mobility behaviour, but both focus on private vehicles and only smart charging instead of V2G.  \cite{gschwendtner2023impact} and \cite{egbue2020multi} propose an agent-based simulation to realistically model (future) charging behavior, while \cite{liu2022spatial} use a similar approach to model EV and V2G adoption behavior. A review of methods to model EV usage is given by~\cite{daina2017modeling}.

Although the possibility of applying V2G on car sharing fleets has been discussed as a business model~\cite{fournier2014carsharing, barahona2023providing}, only few case studies were conducted on real car sharing data~\cite{zhang_values_2020, caggiani_static_2021, xu_electric_2021, prencipe2022mathematical} or data from other commercial vehicle fleets~\cite{zhao2015hybrid, figgener2022influence}. Similarly, there are hardly any future projections of how profitable V2G could be for a car sharing service, with few exceptions that focus on very specific cases such as autonomous EV fleets~\cite{liao2021shared}. We fill this gap with an analysis of a large car sharing fleet in future scenarios.

\section{Methods}\label{sec:methods}

\begin{figure}[ht]
    \centering
    \includegraphics[width=\textwidth]{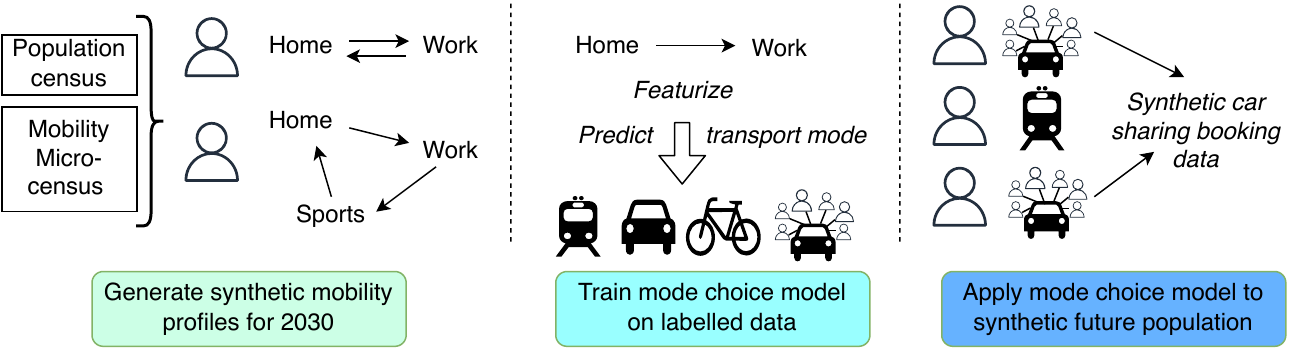}
    \caption{Overview of synthetic car sharing data generation pipeline. A synthetic population with mobility profiles is generated from projected census data. On the other hand, a mode choice model is trained on labeled tracking data, using features such as trip distance, day times, and car availability as input. This model is applied to the synthetic data, yielding a set of synthetic car sharing trips.}
    \label{fig:wp2_overview}
\end{figure}

We simulate car sharing usage for different scenarios for 2030, varying the number of users, the station layout, and the vehicle fleet composition. 
An overview of our pipeline for an agent-based simulation of car reservations is shown in \autoref{fig:wp2_overview}. We propose 1) to generate a future population and their daily activities based on the national Mobility Microcensus, 2) to train a mode choice model using a data-driven approach based on labeled tracking data, and 3) to sample car sharing users from the future population and model their mode choices with the pre-trained model. Furthermore, we simulate the different future scenarios of car sharing services by varying the size of the fleet and stations. Our car sharing simulator is publicly available at \url{https://github.com/mie-lab/car_sharing_simulator}. Each step will be described in the following sections after introducing the main dataset used in the study.

\subsection{Car sharing dataset}

We utilize a national-scale station-based car sharing service in Switzerland. The car sharing service provider, \textit{Mobility}, currently has around 3000 vehicles deployed at more than 1500 stations\footnote{\url{https://www.mobility.ch/en/mobility-cooperative} (accessed 14.2.2023)}. Their business model is strongly based on station-based return trips (only 0.3\% of the trips are one-way) and their customer base consists of members of the \textit{Mobility} cooperative and regular subscriptions. For the purpose of this study, a dataset of all vehicle reservations for the whole year 2019 was used. 
Overall, the dataset comprises 1,284,753 bookings, but only 3.5\% of those were done with electric vehicles. However, the company aims to electrify its whole fleet by 2030. In addition to the reservation data, we also utilize information about the location and size of the car sharing stations, the user age group, gender and approximate living locations (within a 200m radius of the real residence), and vehicle types (combi, budget, transporter, etc) in this study.

\subsection{Simulating activity patterns of a population in 2030}

We utilize previous work in transport planning for synthesizing a future population and their mobility behavior. \cite{horl2021synthetic} presented a pipeline for deriving synthetic populations from census data. It has been applied to regions such as Switzerland, Michigan~\cite{kagho2021demand}, and Sao Paulo~\cite{sallard2021open}; usually for the purpose of generating input data for a MATSim~\cite{w2016multi} simulation. Here, an implementation of the pipeline for Switzerland is used\footnote{Open-source code is available at \url{https://gitlab.ethz.ch/ivt-vpl/populations/ch-zh-synpop\#raw-data}}~\cite{tchervenkov2022switzerland}. The pipeline takes census data as input and outputs daily trips and activities of a synthetic population. 
In detail, the pipeline involves the following stages for our use case (see \cite{tchervenkov2022switzerland} for methods):
\begin{enumerate}
    \item Projecting population data to 2030 via iterative proportional fitting (IPF)~\cite{fienberg1983iterative, choupani2016population}. Population census data from the past five years~\cite{Office2021_STATPOP} are combined with official projections from the Federal Office of Statistics in Switzerland.
    \item Subsampling the synthetic population.
    \item Assigning samples from the Mobility Microcensus~\cite{FOSD2017} (more than 57'000 households) to the synthetic population by statistical matching.
    \item Assigning trips, activities and their spatial locations by sampling from the Mobility Microcensus~\cite{FOSD2017}.
\end{enumerate}

For steps 1, 3, and 4 we apply the implementation by \cite{tchervenkov2022switzerland} without any modifications. However, rather than generating a \textit{random} subset of the population (step 2), our approach focuses on sampling the entire population of car sharing users who hold a car sharing subscription.  As we are not reliant on MATSim but have developed our own agent-based simulation exclusively modeling car sharing, the synthetic population should comprise only those individuals who are potential car sharing users. In other words, simplifying the supply side in our simulation allows us to model a real-sized car sharing user population. 

Since there is no information in the census data regarding car sharing subscriptions, we design a stratified sampling strategy to sample car sharing users. To determine this selection, we take into account the characteristics of current car sharing users in Switzerland, using the dataset from Mobility. In the following, we will refer to the car sharing users in the real dataset as "\textit{Mobility} users", short $U_{real}$, to distinguish them from the synthetic car sharing users $U_{syn}$ that we aim to sample. Let $Q_{syn}$ be the synthetic population generated in step 1, and let $Q_{real}$ be the Swiss population in 2019. To create a subsample of $Q_{syn}$ that resembles car sharing users, we propose to control for age, gender, and proximity to a car sharing station.\footnote{Further relevant characteristics, such as income, are not available for the \textit{Mobility} users.}

For any person $x$, let $a(x)$ be the age of $x$, $g(x)$ its gender, and $s(x)$ the car sharing station that is closest to the place of residence of $x$ ($x\in Q_{syn}$ or $x\in Q_{real}$ or $x\in U_{real}$). The synthetic population is grouped into six age groups to match the age attribute in the \textit{Mobility} dataset. The sample probabilities are derived from the prevalence of age group, gender, and closest station among the Mobility users $U_{real}$, compared to the prevalence in $Q_{syn}$ and $Q_{real}$. The sample weight $w_q$ is set to 
\begin{gather}
\forall q\in Q_{syn}:\ \ 
    w_q = \frac{\sum_{u\in U_{real}} \mathbbm{1}[a(u) = a(q)]}{\sum_{q'\in Q_{real}} \mathbbm{1}[a(q') = a(q)]} \cdot
    \frac{\sum_{u\in U_{real}} \mathbbm{1}[g(u) = g(q)]}{\sum_{q'\in Q_{real}} \mathbbm{1}[g(q') = g(q)]} \cdot
    \frac{\sum_{u\in U_{real}} \mathbbm{1}[s(u) = s(q)]}{\sum_{q'\in Q_{real}} \mathbbm{1}[s(q') = s(q)]}
\end{gather}
%
Intuitively, person $q$ is sampled with higher probability (larger $w_q$) if its age, gender, and the closest car sharing station are more prevalent among car sharing users $U_{real}$ than among the general population $Q_{real}$. By normalizing with the prevalence in $Q_{real}$ instead of $Q_{syn}$, the sociodemographics of $U_{syn}$ can differ from the ones of $U_{real}$ (the current car sharing users), subject to changes in general population ($Q_{syn}$ compared to $Q_{real}$).\\
The weights $w_q$ are normalized to ensure $\sum_{q\in Q_{syn}} \hat{w_q} = 1$ and a set of $N$ car sharing users, $U_{syn}$, is sampled from $Q_{syn}$ with $\hat{w_q}$ as the sample probabilities. $N$ is a parameter that is varied by scenario. As a result of the stratified sampling, the age, gender, and distance to the closest car sharing station of $U_{syn}$ are similarly distributed as for the actual \textit{Mobility} customers $U_{real}$ (see Appendix~\ref{app:carsharinguser_sampling}). 

After generating the synthetic population $U_{syn}$, executing step 3 and 4 of the pipeline by \cite{tchervenkov2022switzerland} yields the activity profiles for all $u\in U_{syn}$ for \emph{a single day}. An activity profile is a sequence of activities and their spatial locations (see \autoref{fig:wp2_overview}), including start and end time, reflecting a person's activity schedule of one day. All activities have a start time, a purpose (home / leisure / work / shopping / education / other), and a location (spatial coordinates). For each user, we convert the activities of a user into \textit{trips} between activities. All trips have origin and destination locations, origin and destination purpose, and origin and destination start time. $m$ activities for one user result in $m-1$ trips, where the transport mode is unknown.

\subsection{Data-driven mode choice modeling for car sharing usage}\label{sec:modechoicemodel}

Numerous factors influence an individual's choice of transportation mode, including trip distance, activity type, pricing, travel duration, and proximity to public transport or car sharing. We propose a machine learning model to capture the complex interplay of these features with a data-driven approach. This model necessitates a labeled dataset in which the mode of transport is known, and it must include car sharing trips. We found the MOBIS dataset~\cite{molloy2022mobis} to be highly suitable. MOBIS, a GPS tracking study conducted in Switzerland in 2019, aimed to analyze nudging and pricing incentives for mode choices. After the initial 14-week tracking period with 3680 participants, the users were asked to re-activate the app voluntarily to study travel behavior changes during the COVID-19 pandemic~\cite{molloy2022mobiscovid, heimgartner2022understanding}. Users were further asked to manually label the transport mode and activity purpose. As a result, the MOBIS-COVID studies offer a rich labelled dataset with tracking data over more than one year. Crucially, the trips in the MOBIS dataset include \textit{Mobility} car sharing as a transport mode. We found 225 users with at least one car sharing trip. In the following, we restrict ourselves to these users, as our objective is to learn the likelihood of an agent booking a shared vehicle when a car sharing subscription is in place. The 225 car sharing users recorded 346712 triplegs\footnote{For definition of triplegs, please refer to \cite{martin2023trackintel}.} in total within the time period from September 2019 until January 2022, and 2270 (0.65\%) were labeled as car sharing. 

The MOBIS data are provided as \textit{activities} and \textit{triplegs} that were derived from raw GPS data. We first aggregate the triplegs into trips with the Trackintel library~\cite{martin2023trackintel}, following the data model described by~\cite{schonfelder2016urban}. For example, two consecutive triplegs that are interrupted only by a negligible activity, such as waiting for a bus, are merged into one trip. This aggregation step yields 102282 trips. Since the triplegs comprising one trip may have different modes of transport, we set the trip-wise transport mode to the one covering the largest distance of the trip. For each trip, we extracted a set of features that were determined based on availability (for both MOBIS and synthetic data) and relevance to transport mode prediction. A full list of features is shown in \autoref{tab:feat_comp}, and includes socio-demographics of the user (age, gender), trip distance, activity features (start time and purpose of current and next activity), and accessibility features (public transport accessibility score, distance to closest car sharing station).

We aim to model mode choices of car sharing users by training a machine learning model to predict the transport mode, given the trip features as input. The model is trained on the MOBIS data but will be applied to assign a mode to trips of the synthetic population.  Therefore, we first investigate whether the MOBIS data are sufficiently similar to the synthetic trip data. \autoref{tab:feat_comp} demonstrates that 
the distribution of feature values of the synthetic trips aligns well with the distribution in the MOBIS data, at least in terms of the first two moments. The difference is also statistically estimated by means of z-scores, calculated as the difference of the mean feature value divided by the standard deviation within the MOBIS dataset\footnote{z-scores were computed as (mean(synthetic\_population) - mean(MOBIS)) / std(MOBIS)}. The z-scores are all below 0.6 in absolute value, confirming the similarity in trip characteristics.

\begin{table}[ht]
\centering
\resizebox{0.85\textwidth}{!}{
\begin{tabular}{lrrr}
\toprule
{} &               
MOBIS car sharing users
& Synthetic car sharing population &  z-value\\
Feature                         &      mean (std)               &       mean (std)               &                                         \\
\midrule
user age                             &       49.05 (13.24) &        42.81 (13.33) &                                   -0.47 \\
user gender                             &         0.39 (0.49) &          0.38 (0.49) &                                   -0.03 \\
user has car access                      &          0.85 (0.3) &          0.78 (0.37) &                                   -0.24 \\
user is employed                        &         0.84 (0.37) &          0.83 (0.37) &                                   -0.01 \\
half-fare public transport subscription                         &         0.58 (0.49) &          0.37 (0.48) &                                   -0.42 \\
full-fare public transport subscription                              &         0.19 (0.39) &          0.08 (0.28) &                                   -0.26 \\
distance from origin to destination                     &  8725.35 (17558.07) &    6166.87 (14247.3) &                                   -0.15 \\
purpose destination = home        &          0.4 (0.49) &           0.4 (0.49) &                                   -0.01 \\
purpose destination = leisure     &         0.22 (0.41) &          0.15 (0.36) &                                   -0.16 \\
purpose destination = work        &         0.23 (0.42) &          0.24 (0.43) &                                    0.03 \\
purpose destination = shopping    &         0.07 (0.26) &            0.1 (0.3) &                                    0.11 \\
purpose destination = education   &          0.01 (0.1) &          0.01 (0.11) &                                    0.02 \\
purpose origin = home             &          0.4 (0.49) &           0.4 (0.49) &                                    0.00 \\
purpose origin = leisure          &         0.22 (0.41) &          0.15 (0.36) &                                   -0.16 \\
purpose origin = work             &         0.23 (0.42) &          0.24 (0.43) &                                    0.02 \\
purpose origin = shopping         &         0.08 (0.26) &            0.1 (0.3) &                                    0.11 \\
purpose origin = education        &          0.01 (0.1) &          0.01 (0.11) &                                    0.02 \\
PT accessibility (origin)       &         2.11 (1.44) &           2.7 (1.33) &                                    0.40 \\
PT accessibility (destination)  &         2.11 (1.44) &           2.7 (1.33) &                                    0.41 \\
distance to station (origin)      &   1426.53 (2403.34) &     950.65 (1735.61) &                                   -0.20 \\
distance to station (destination) &   1430.92 (2405.02) &     950.68 (1735.83) &                                   -0.20 \\
start hour of activity at origin           &        13.88 (4.57) &         14.25 (5.03) &                                    0.08 \\
day of activity at origin                       &          2.84 (1.9) &          1.76 (0.43) &                                   -0.57 \\
start hour of activity at destination                &        13.93 (4.56) &         13.58 (4.82) &                                   -0.08 \\
day of activity at destination                   &         2.85 (1.89) &          1.77 (0.45) &                                   -0.57 \\
\bottomrule
\end{tabular}
}
\caption{Comparing characteristics of trips in the MOBIS dataset (car sharing users) to the synthetically generated trips. The z-value indicates the difference between the respective mean of the feature value, normalized by the standard deviation among the synthetic population. As desired, there are no significant differences.}
\label{tab:feat_comp}
\end{table}

\begin{figure}[ht]
    \centering
    \includegraphics[width=0.95\textwidth]{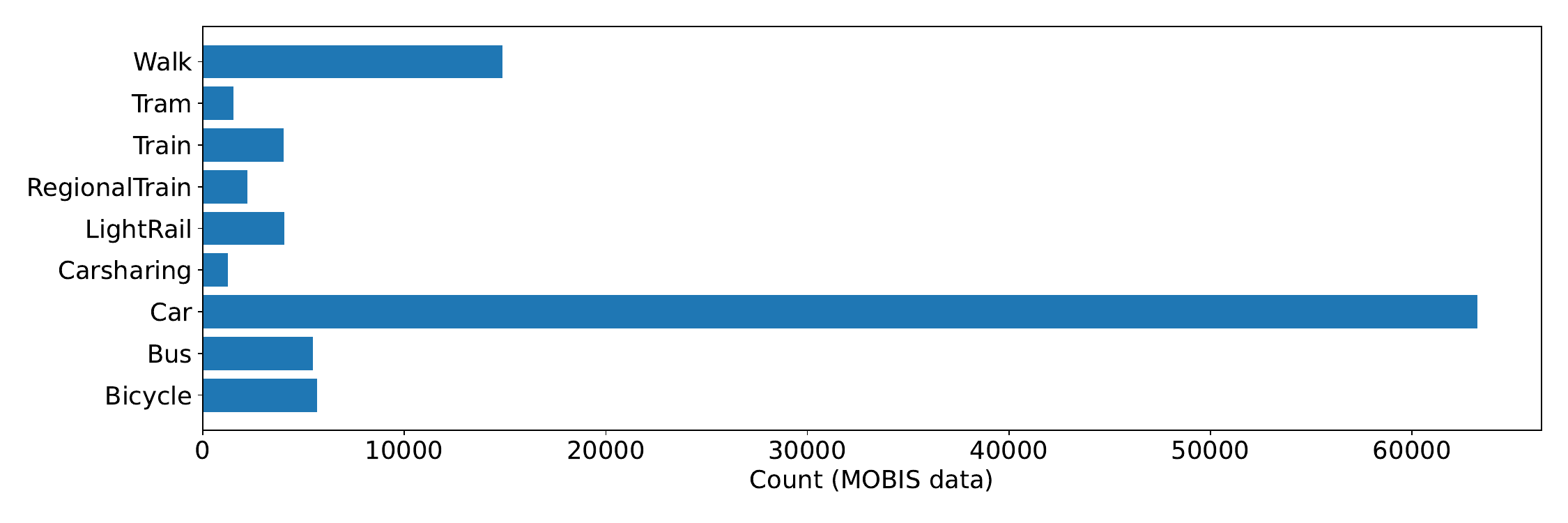}
    \caption{Mode share among trips of MOBIS users with a car sharing subscription}
    \label{fig:mobis_mode_share}
\end{figure}

Recently, supervised ML approaches such as (Gradient Boosted) Decision Trees or Artificial Neural Networks were increasingly used for modeling mode choices~\cite{hillel2021systematic}, in contrast to previously popular Random Utility Models~\cite{mcfadden1981econometric}. Our task can be framed as a multi-class classification with unbalanced class prevalence since the modes "car" or "walking" appear more often than other modes. Only modes with at least 500 occurrences in the dataset are included, excluding rare modes such as boats or cable cars. The final selection of modes and their distribution is shown in \autoref{fig:mobis_mode_share}. A model that was shown to be well-suited for unbalanced datasets is XGBoost (XGB)~\cite{chen2016xgboost}, outperforming other methods in previous work~\cite{wang2018machine}. We, therefore, train an XGB classifier from the \texttt{xgboost} Python package\footnote{\url{https://xgboost.readthedocs.io/en/stable/python/index.html}} and tune the \texttt{max\_depth} parameter with grid search on a validation set from the MOBIS dataset. The model is denoted as $M$ in the following. 

Additionally, we tested an Inverse Reinforcement Learning (IRL) approach. IRL is suitable for problems where the reward function is unknown, and one can only learn to imitate certain behaviors by observing expert demonstrations. Since the objective functions of humans are usually unknown, IRL was used for modeling mobility patterns. \cite{pang2017modeling}, for example, use the maximum entropy IRL approach~\cite{aghasadeghi2011maximum} to generate realistic human trajectories. Here, the task of mode choice prediction can be modeled as an IRL problem where the states correspond to the features of an upcoming trip and the actions correspond to mode choices. Since the states are continuous and the reward function is non-linear, we employ an IRL approach named Guided Cost Learning~\cite{finn2016guided}, where the cost function is learned with a Neural Network. The cost function and the agent policy are trained in alternating fashion. For details on the training, we refer to \cite{finn2016guided} and the open-source implementation\footnote{\url{https://github.com/nishantkr18/guided-cost-learning}}.

\subsection{Simulating car sharing users' reservations}\label{sec:methods_simulation}

The trained mode choice model $M$ is then applied to the trips of the synthetic population. Since there is no explicit information in either the MOBIS data or the synthetic data regarding \textit{when} the mode is chosen, we assume that all users decide spontaneously prior to the upcoming activity, with an adequate buffer. Specifically, we calculate the required travel duration from the current to the next activity assuming 50km/h speed, and add an extra 10 minutes as buffer. For instance, if the next activity is 25km away and is scheduled to begin at 3 pm, the ''mode decision time'' $t_{decision}$ would be set to 2:20 pm. 

A formal explanation of this process is given in \autoref{alg:simulate}. The trips, denoted as $p_1, \dots, p_n$, are considered in the order of their respective decision time $t_{decision}(p)$. At the designated decision time of a trip $p$, the distance to the nearest car sharing station \textit{with an available vehicle} is computed. Together with other features of $p$, this distance serves as an input to the mode choice model $M$ (see \autoref{tab:feat_comp}). $M$ is then employed to predict the mode of $p$ and, whenever $M$ predicts car sharing as the chosen transport mode, a shared vehicle is assigned to the user of the respective trip. 

Since the \textit{Mobility} car sharing service only offers \textit{return} trips, our simulation is designed so that the synthetic users keep the shared car until they return to the location where the car was borrowed. As a result, if a user was assigned a shared vehicle for the previous trip and it was not returned, the shared vehicle is simply reassigned for the next trip, and this process continues until the user re-visits the location where the car was borrowed for another activity. An alternative approach to modeling return trips involves introducing new trips into the user's mobility profiles to return the car after a certain duration (e.g. sampled durations). However, we opt for strictly following the synthetic mobility profiles to avoid deviating from the agent-based viewpoint and the given synthetic data. In future work, this challenge can be approached by learning the mode choice for a complete sequence of activities, in order to account for mode dependencies between trips.

The output of the simulation (\autoref{alg:simulate}) is a set of car sharing reservations $R$. They are merged temporally if the same vehicle was assigned for multiple sequential trips of the same person. $R$ is a dataset similar to the available \textit{Mobility} data, listing vehicle reservations with start and end time, driven kilometers, duration, and user ID.

\begin{algorithm}[tb]
   \caption{Simulate car sharing reservations}
   \label{alg:simulate}
    \begin{algorithmic}
       \STATE \textbf{Input}: Mode choice model $M$\\
       \STATE \textbf{Input}: Planned trips $p_1, \dots p_n$ (without mode of transport) \\
       \STATE \textbf{Input}: Trip information: user $u(p_i)$, origin and destination location $l_{origin}(p_i), l_{destination}(p_i)$, start time of the activity at the destination $t_{destination}(p_i)$, and distance $d_{trip}(p_i)$ 
       \vspace{1.5em}
        \STATE \textbf{Initialize:}
        \STATE $mode(p_i) = unknown\ \forall p_i$
       \STATE $ \text{car\_reserved}(u) = 0\ \  \forall u$ \tcp*{Initialize all users to have no car reserved}
        \STATE $t_{decision}(p_i) = t_{destination}(p_i) - \frac{d_{trip}(p_i)}{1000 \cdot 50 km/h} \cdot 60 - 10\text{min}$ \tcp*{Compute mode decision time}
        \STATE $P = sorted(p_1, \dots, p_n)$ \tcp*{Sort trips by $t_{decision}(p_i)$}
        \STATE $R = \{ \}$ \tcp*{Initialize car sharing reservations}
        \FOR{$p_i\in P$}
            \IF{$\text{car\_reserved}(u(p_i)) = 0$ \tcp*{Option 1: No car currently - predict transport mode with M}} 
                 \STATE Compute the distance to the closest station with a free vehicle $d_{station}(p_i)$\\
                \STATE $mode(p_i) = M\big(d_{trip}(p_i), d_{station}(p_i), \text{other trip features} \big)$ \tcp*{Apply mode choice model}
                \IF{$mode(p_i) = shared$}
                    \STATE $\text{car\_reserved}(u(p_i)) = 1$ \tcp*{Save whether the user has a shared car}
                    \STATE $R = R \cup p_i$ \tcp*{Add trip to car sharing bookings}
                \ENDIF
            \ELSE 
                \STATE $mode(p_i) = shared$  \tcp*{ Option 2: The user still has a shared car -> keep it }
                \STATE $\text{car\_reserved}(u(p_i)) = 1$
                \IF{$l_{destination}(p_i)$ is equal to start location of reservation}
                    \STATE Return car to station at time $t_{destination}(p_i)$
                    \STATE $\text{car\_reserved}(u(p_i)) = 0$
                \ENDIF
            \ENDIF
        \ENDFOR
        \RETURN $R$
    \end{algorithmic}
\end{algorithm}

\subsection{Simulating the car sharing service}\label{sec:station_placement}

Following the announced goal of \textit{Mobility}, we assume a fully electric fleet by 2030. In consultation with the company, we assign one EV model to each vehicle category, e.g. the e-up model by Volkswagen for the "Budget" category, the Tesla Model 3 for the "Premium" category, the eVito by Mercedes Benz for the "Transporter" category, etc. While this does not reflect the potential vehicle diversity in the future fleet, it simplifies the modeling of user behaviors. 

Second, we generate scenarios with a larger number of vehicles per station, to account for a potential increase of supply at each station. Given a desired fleet size $V_{desired}$, e.g. 3000 vehicles, the number of vehicles per station is scaled accordingly by a factor of $c= \frac{V_{desired}}{V_{current}}$. Due to the discrete nature of the vehicle count per station, we scale the station-wise vehicle count by a factor $\hat{c}$ that is normally distributed around $c$, $\hat{c} \sim \mathcal{N}(c, 0.3)$, and round the result, such that the expected number of vehicles is the desired number of vehicles, $\mathbb{E}[V] = V_{desired}$. We redo this sampling process until the deviation from $V_{desired}$ is lower than 0.5\%. The new vehicles are assigned to a category of car sharing vehicles (e.g., Budget, Premium, Transporter, etc.) by sampling from a categorical distribution with probabilities corresponding to the current share of the respective category in the \textit{Mobility} dataset.

Finally, we propose an algorithm to simulate the placement of new stations, which is crucial for scenarios that assume an expansion of the car sharing service. Since simulating a realistic allocation of new stations is out of scope for this study, we adopted a simplistic model that assumes new stations are placed based on the population density while avoiding existing stations. This scenario bears a resemblance to the well-known KMeans clustering problem~\cite{macqueen1967some}, where the goal is to minimize the distance of samples (i.e. potential customers) to the cluster centers (i.e. car sharing stations). The KMeans objective function is NP-hard, but an iterative algorithm by \cite{lloyd1982least} generally yields good solutions. In contrast to the KMeans clustering problem, our situation assumes that some of the cluster centers remain \textit{fixed}, corresponding to the existing stations. Thus, we adapt Lloyds algorithm to update only a subset of the cluster centroids during each iteration, preserving the existing station locations. In detail, we sample 500k residential locations from the synthetic population, denoted as $X$, to represent the spatial distribution of potential customers. The new stations, $\mu_0, \dots, \mu_k$, are initially randomly placed by sampling from $X$ and are subsequently relocated iteratively to areas with higher population density and minimal proximity to existing car sharing stations. The full algorithm is provided in \autoref{alg:kmeans}.

\begin{algorithm}[htb]
   \caption{KMeans for new station placement}
   \label{alg:kmeans}
    \begin{algorithmic}
       \STATE \textbf{Input}: Customer home locations $X = \{x_1, x_2, \dots, x_n\}$
       \STATE \textbf{Input}: Number of desired new stations $k$
       \STATE \textbf{Input}: Set of fixed station locations $S_{fix}$
       \STATE $t = 0$
       \STATE $C = S_{fix} \cup \{\mu_1^0, \mu_2^0, \dots, \mu_k^0\}$ \tcp*{Initialize cluster centroids at $t = 0$ (sample $\mu^0$ from X)}
       \REPEAT
       \STATE Assign each data point $x_i$ to the closest cluster centroid from $C$
       \STATE $\mu_i^{t+1} = \frac{1}{n} \sum_i^n x_i \cdot \mathbbm{1}[x_i\text{ was assigned to } \mu_i^t]$ \tcp*{Update non-fixed cluster centroids}
       \STATE $C = S_{fix} \cup \{\mu_1^{t+1}, \mu_2^{t+1}, \dots, \mu_k^{t+1}\}$
       \tcp*{Update full set of cluster centroids}
       \STATE $t= t+1$
       \UNTIL{convergence}
    \end{algorithmic}
\end{algorithm}

\begin{figure}[htb]
    \centering
    \includegraphics[width=\textwidth]{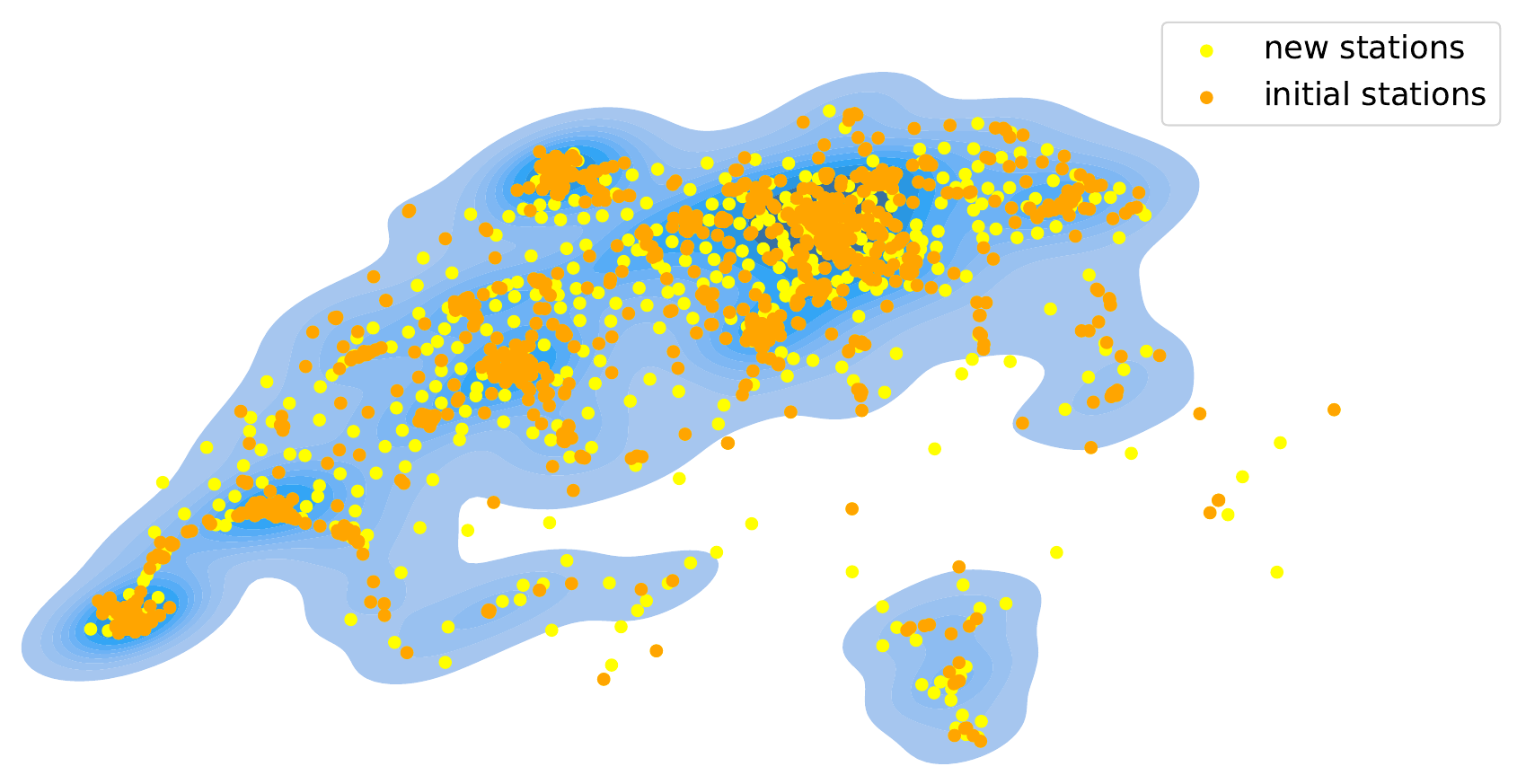}
    \caption{Exemplary distribution of new stations in the simulated car sharing service. For visualization purposes, the population locations are displayed as a distribution (blue) via Kernel Density Estimation. 
    The new stations generated with our algorithm (yellow) follow the distribution of former stations (orange) and cover additional regions with high population density.}
    \label{fig:station_placing}
\end{figure}

\autoref{fig:station_placing} depicts an illustrative output obtained by applying \autoref{alg:kmeans} to deploy 1000 new stations in Switzerland. The population density is shown in blue. As intended, the newly placed stations align with the population density, effectively bridging gaps in the distribution of existing stations.

\subsection{Simulating V2G}

Finally, we simulate V2G operations using an optimization approach. While substantial research exists on optimizing charging and discharging of vehicles, even in car sharing contexts~\cite{zhang_values_2020, caggiani_static_2021, xu_electric_2021}, hardly any approach scales to a large, national-scale car sharing fleet, as in our scenarios. To address this issue, \cite{nespoli2022national} have recently developed a scheduling approach for EV fleet control that scales to thousands of vehicles, leveraging the Alternating Direction Methods of Multipliers (ADMM). The method allows the decomposition of the problem by car sharing station while still optimizing a fleet-level objective. For details on the mathematical formulation, see \cite{nespoli2022national}.



\section{Results}\label{sec:res}
To validate our simulation, we first compare a simulated dataset to the real car sharing dataset from 2019. We then generate scenarios for 2030 and discuss the resulting opportunities for V2G.

\subsection{Validation}


\subsubsection{Performance of the mode choice model}

\begin{figure}
    \centering
    \begin{subfigure}[b]{0.8\textwidth}
    \includegraphics[width=\textwidth]{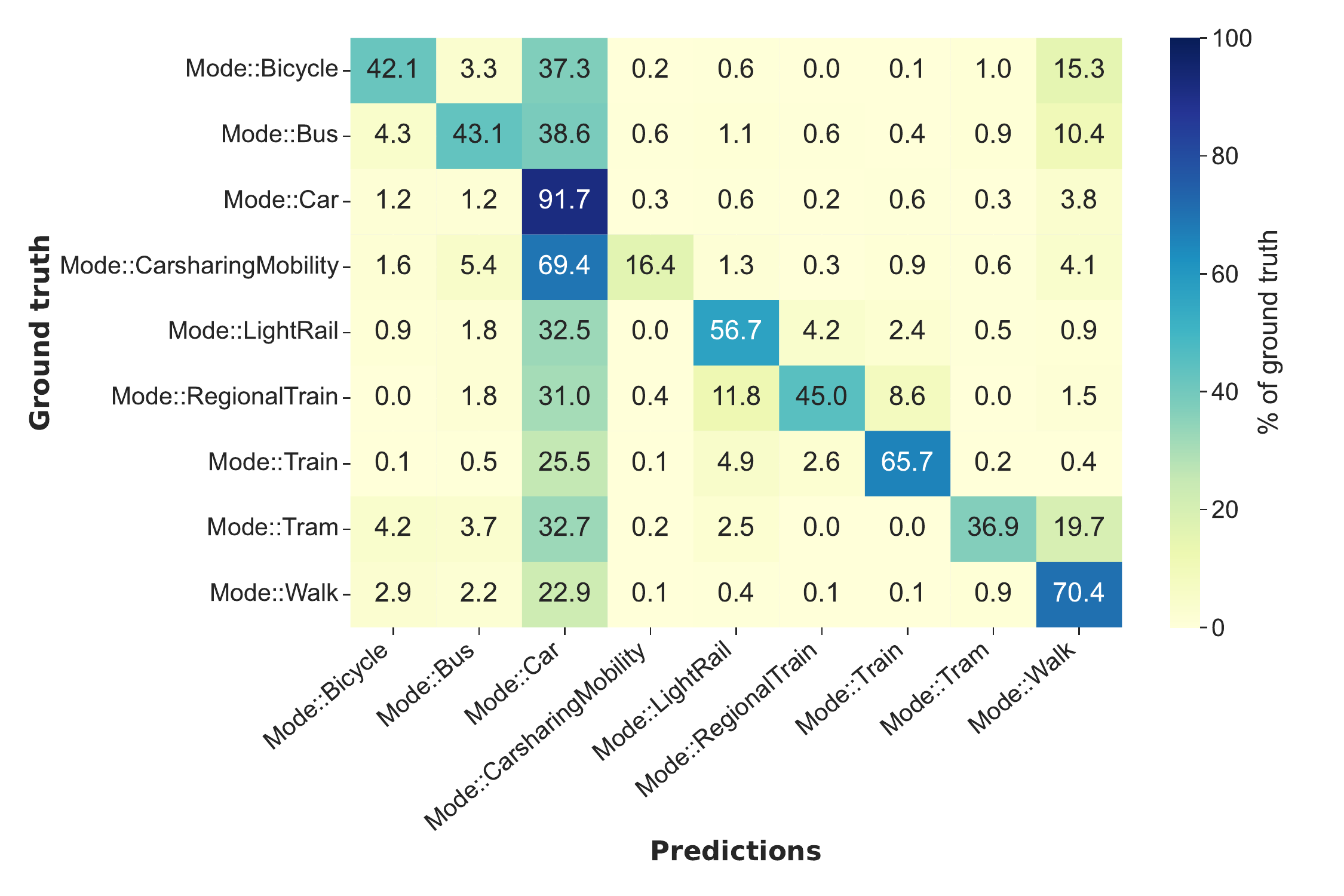}
    \caption{Normalized by true mode}
    \label{fig:bytrue}
    \end{subfigure}
    \hfill
    \begin{subfigure}[b]{0.8\textwidth}
    \includegraphics[width=\textwidth]{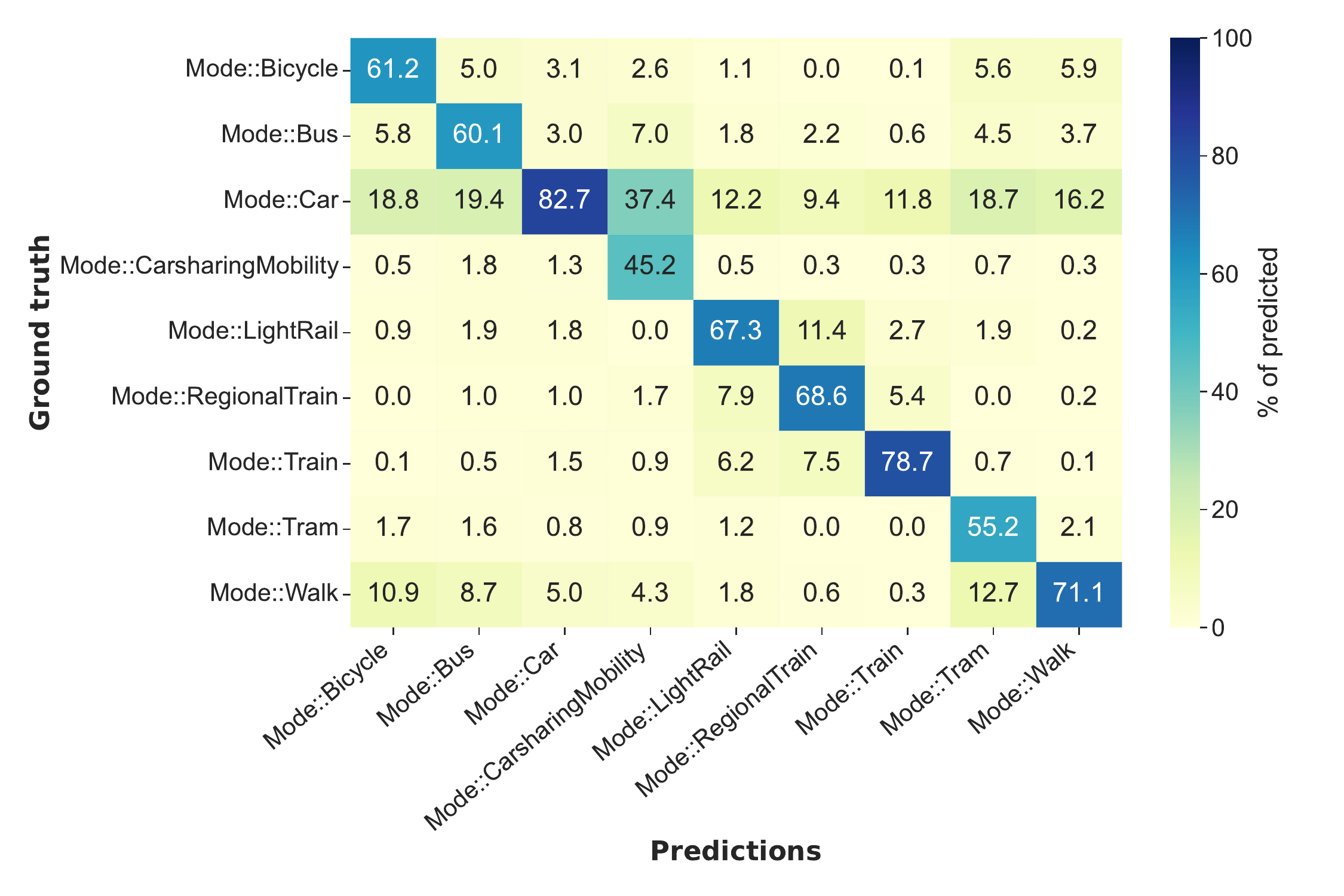}
    \caption{Normalized by predicted mode}
    \label{fig:bypred}
    \end{subfigure}
    \caption{Confusion matrix of mode choice model on test set.}
    \label{fig:confusionmatrix}
\end{figure}

The training accuracy of the XGB model (see \autoref{sec:modechoicemodel}) is 81.7\% (80.0\% balanced accuracy over classes). When tested on a separate hold-out dataset from MOBIS, the model achieved an accuracy of 77.9\% (65\% balanced accuracy). \autoref{fig:confusionmatrix} shows the confusion matrices normalized by ground truth and by the predicted label. 
Notably, we achieved a specificity of 45\% in predicting car sharing trips, a high value considering the similarities car sharing shares with conventional car trips.
However, the sensitivity for car sharing predictions was lower, with only 16\% of actual car sharing trips correctly identified (\autoref{fig:bytrue}). This is partly attributed to the relatively low prevalence of car sharing trips within the broader transportation options landscape, as illustrated in \autoref{fig:mobis_mode_share}, which naturally increases the challenge of identifying car sharing trips.
Nevertheless, for our simulation, what holds paramount is that the model replicates the distribution of travel modes. It is evident that car sharing is mainly confused with car trips, aligning with our expectations, and all travel modes are represented in the predictions. This capability makes our model well-suited for generating realistic mode shares within synthetic populations. When applied to a synthetic population for 2019 (without considering vehicle availability), car sharing has a share of 0.926\%, aligning closely with the share in the MOBIS data (0.965\%). Appendix~\ref{app:mode_choice_stability} further confirms the model's stability over time, reinforcing its suitability for generating projections to 2030.

\begin{figure}
    \centering
    \includegraphics[width=\textwidth]{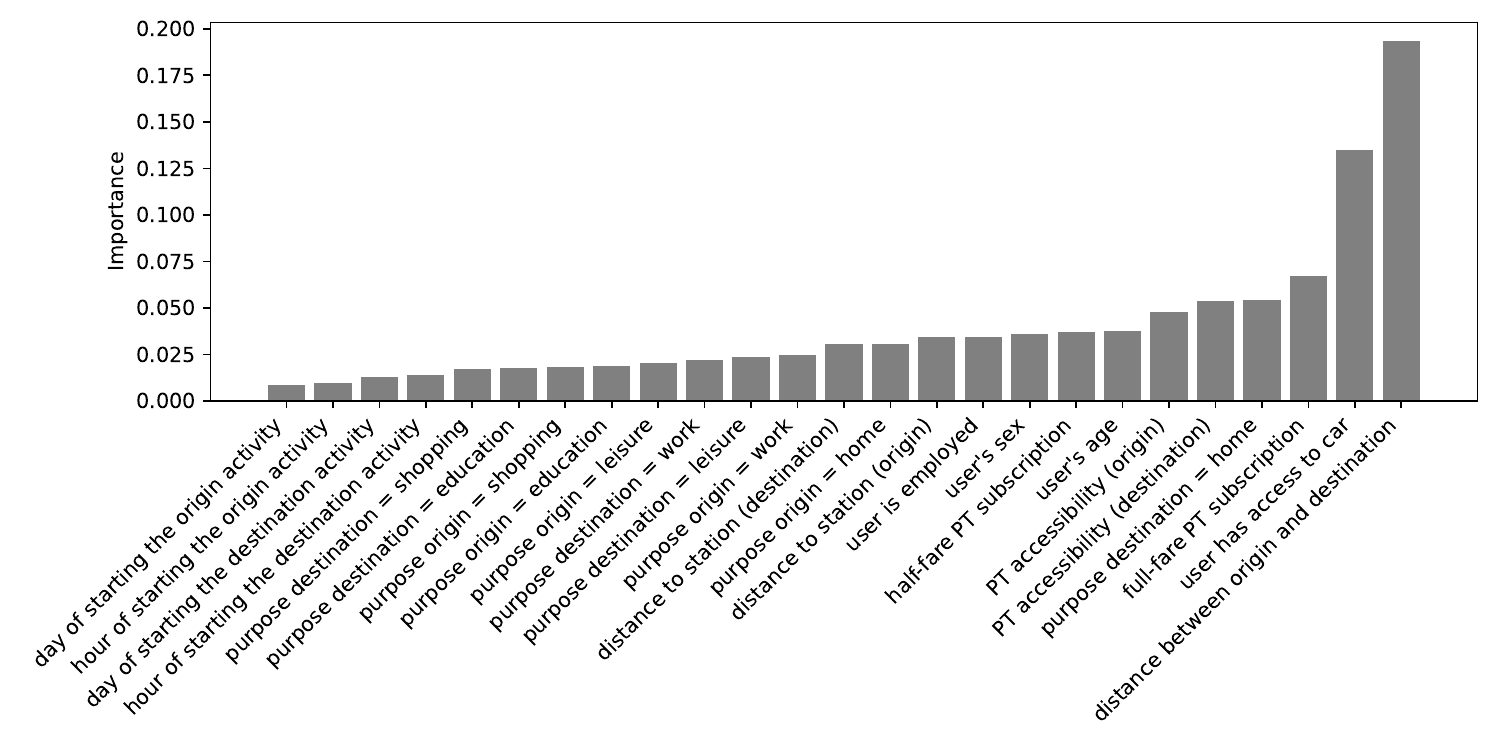}
    \caption{Importance of feature in the XGBoost model. The distance is most important as expected, but also socio-demographic features play an important role.}
    \label{fig:feature_importance}
\end{figure}

Since XGBoost models rely on decision trees as base estimators, the importance of each input feature can be estimated in terms of the relevance of the feature in branching the trees. \autoref{fig:feature_importance} illustrates that the trip distance emerges as the most pivotal feature in our model. User attributes such as car access, the availability of full-fare or half-fare public transport subscriptions in Switzerland, and age exhibit notable relevance. This property is advantageous for simulating future car sharing behavior as it enables the model to reflect sociodemographic changes in a population. 

\subsubsection{Calibration of the agent-based simulator}

To validate our simulation pipeline, we apply it to 2019. Using the pipeline by \cite{tchervenkov2022switzerland}, we generate a synthetic population, apply the pre-trained mode choice models, and derive synthetic car sharing reservations for 2019. The car sharing behavior is then compared to the real car sharing dataset from 2019. If our simulation is realistic, we would expect that the reservation variables are similar to the real data \textit{in distribution}. In particular, we can compare car reservation start and end times, their distance and duration, and the distribution over stations. Note that the following comparison is between one day of simulated data and a full year of real reservations, since the pipeline by \cite{tchervenkov2022switzerland} only generates activity profiles of a synthetic population for a single day. Bookings longer than one day are excluded from the real data for a fair comparison.

\begin{figure}[htb]
    \centering
    \begin{subfigure}[b]{0.58\textwidth}
    \includegraphics[width=\textwidth]{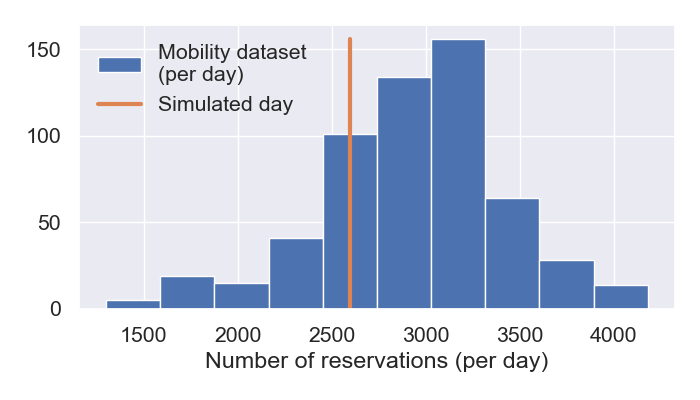}
    \caption{Number of simulated reservations compared to days in the real data}
    \label{fig:nrreservation}
    \end{subfigure}
    \hfill
    \begin{subfigure}[b]{0.4\textwidth}
    \includegraphics[width=\textwidth]{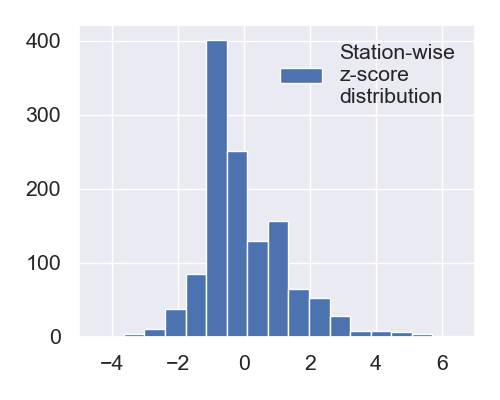}
    \caption{Z-score of reservation count per station}
    \label{fig:zscorestation}
    \end{subfigure}
    \caption{With 100000 car sharing users in the simulated population, the number of simulated car sharing reservations lies within the expected range (\subref{fig:nrreservation}). The z scores over stations show that most stations are used with similar frequency with few exceptions (\subref{fig:zscorestation}).}
    \label{fig:stationdist}
\end{figure}

First, we report the total number of reservations of the simulated day compared to the typical daily reservation count in the \textit{Mobility} dataset in \autoref{fig:nrreservation}. The number of reservations is directly influenced by the number of simulated car sharing users $N$. However, the frequency of car sharing bookings for MOBIS users differs significantly from the \textit{Mobility} users. The \textit{Mobility} dataset comprises 270k users, yet only 117k users appear at least once in the entire year of 2019, indicating a considerable number of inactive users. As a result, it is necessary to calibrate $N$ to attain a reservation count that aligns with reality. By calibration, $N=100000$ was established as a realistic estimate for the number of users who actively consider car sharing in their daily mode choices.

\begin{figure}[htb]
    \centering
    \includegraphics[width=0.6\textwidth]{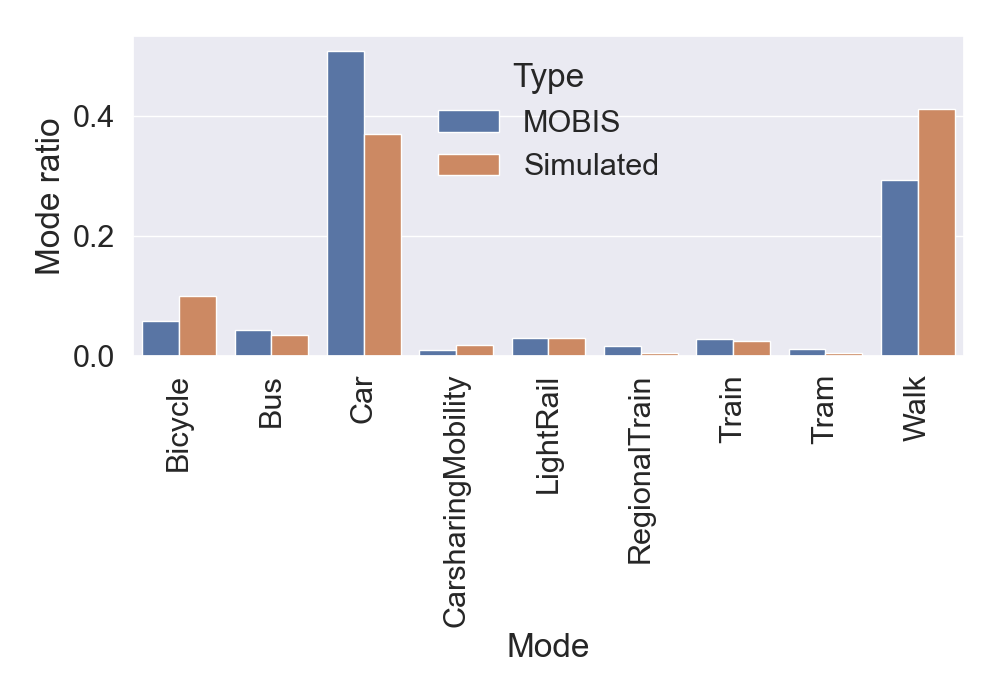}
    \caption{Applying the trained mode choice model to the simulated population yields a similar mode share as for the MOBIS data, apart from a smaller fraction of car trips and a larger fraction of walk trips.}
    \label{fig:modeshare}
\end{figure}

\begin{figure}[bt]
    \centering
    \begin{subfigure}[b]{\textwidth}   
    \includegraphics[width=\textwidth]{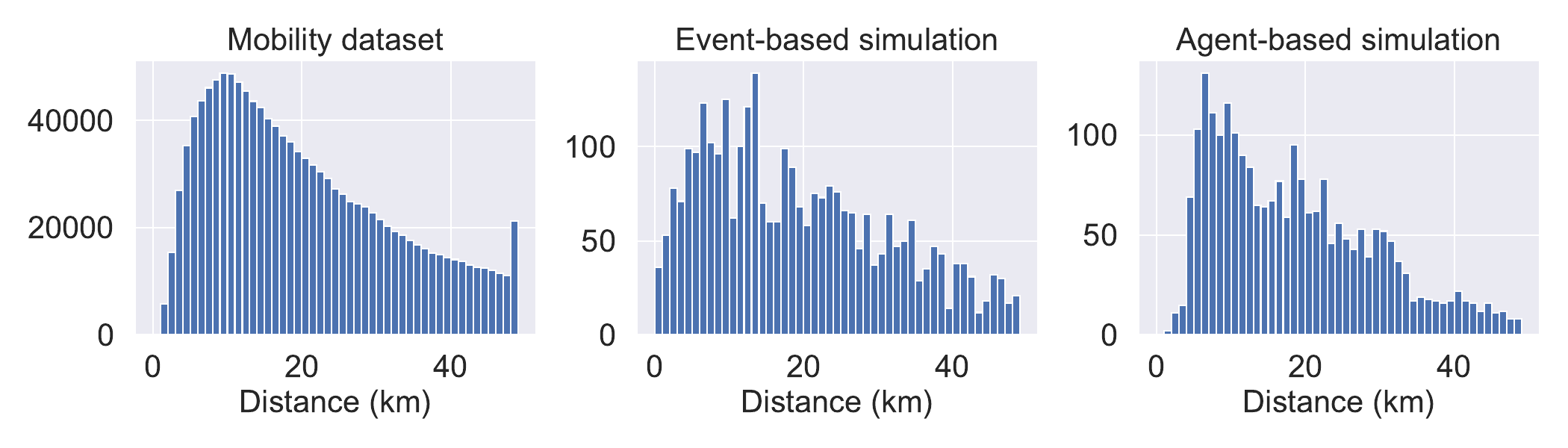}
    \caption{Driven distance}
    \end{subfigure}
    \begin{subfigure}[b]{\textwidth}   
    \includegraphics[width=\textwidth]{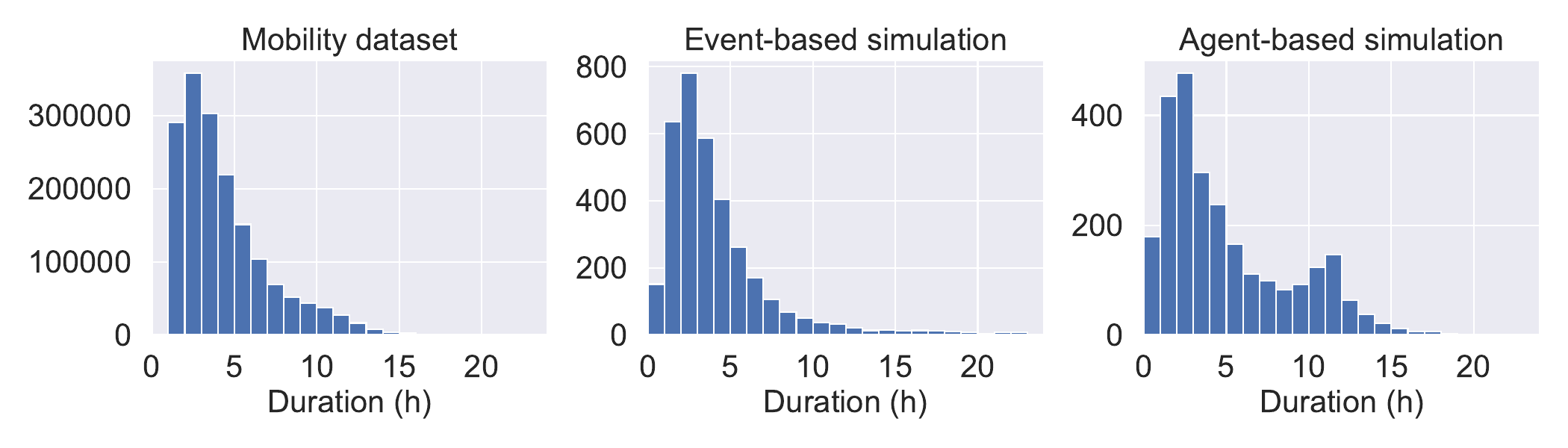}
    \caption{Duration}
    \end{subfigure}
        \begin{subfigure}[b]{\textwidth}   
    \includegraphics[width=\textwidth]{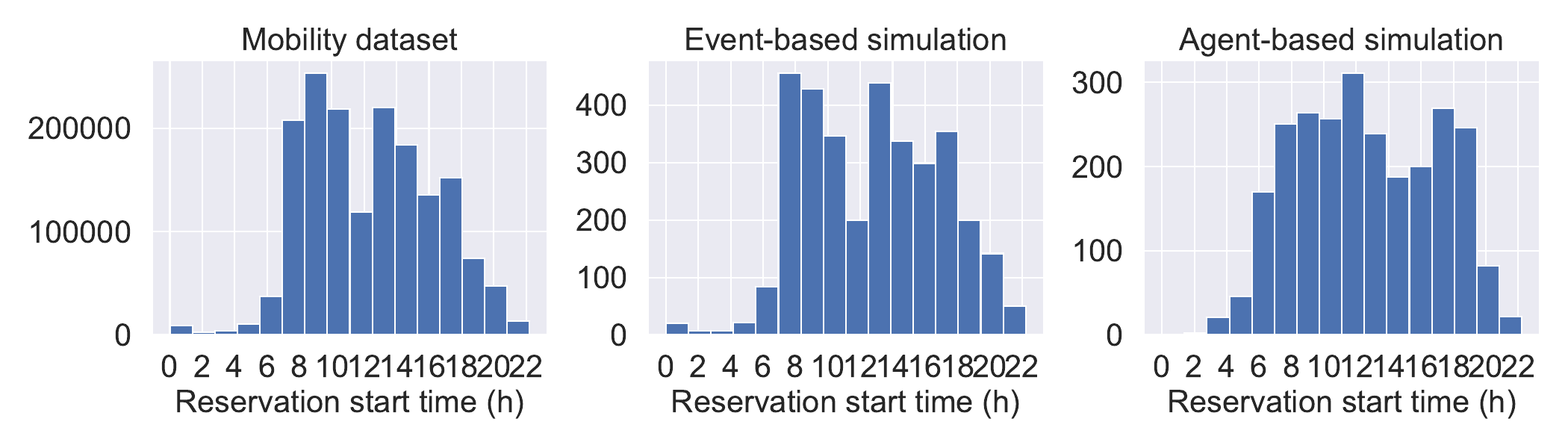}
    \caption{Reservation start time}
    \end{subfigure}
        \begin{subfigure}[b]{\textwidth}   
    \includegraphics[width=\textwidth]{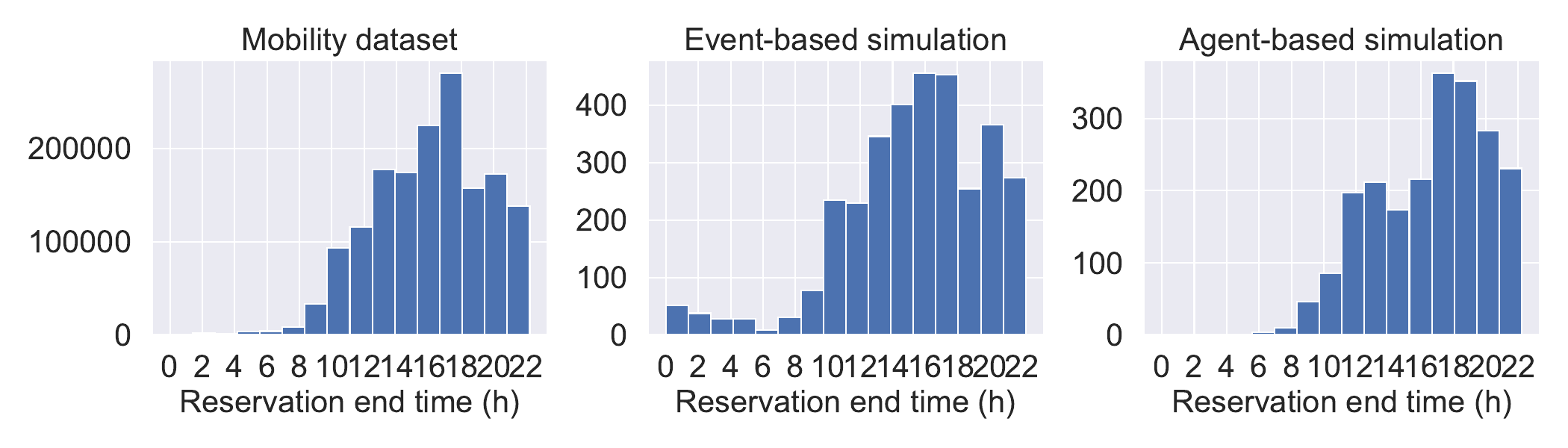}
    \caption{Reservation end time}
    \end{subfigure}
    \caption{Comparing simulated to real data from 2019 in distribution}
    \label{fig:validation_eventbased}
\end{figure}

Secondly, in \autoref{fig:zscorestation} we compare the distribution of bookings \textit{by stations} to the real data in 2019. 
In this case, we compute a z-score for each individual station $s$, since the number of bookings per station is highly variable. Let $\mu_s$ be the real average daily number of reservations for station $s$ and $\sigma_s$ its standard deviation, and let $y_s$ be the number of reservations for $s$ in the simulation. We compute a z-score as $z = (y_s - \mu_s) / \sigma_s$. The distribution of z-scores is shown in \autoref{fig:zscorestation}. As desired, the z-scores are mainly low and are distributed similarly as for an event-based simulation (see Appendix~\ref{app:eventbased}). There is a small negative bias; i.e., there are more stations with a lower number of trips than in the real dataset. This is due to the large number of stations with only one car, which is not used in a single day. 

Third, we validate the mode choice model by comparing the modal share of the simulated population to the modal share of MOBIS users. As \autoref{fig:modeshare} shows, the mode share is very similar, with a slight shift from car trips to bike and walking trips. However, the MOBIS data only includes 225 users that are not representative of the real population.

Last, in \autoref{fig:validation_eventbased} the data generated with our simulation is contrasted to the real data and to data sampled from an event-based simulator, following \cite{cocca_free_2019} (for implementation details, see Appendix~\ref{app:eventbased}). The simulated bookings align well with the real bookings in terms of start and end time distribution, reservation duration, and driving distance in kilometers. The driving distance is underestimated by our agent-based simulation, since the beeline distance between agents' activities is used, which can be further refined by applying map-matching. The distribution of the duration deviates from the real data, with a larger fraction of 5-15 hours car sharing rentals. This can be attributed to our simulation of \textit{return} trips (see \autoref{sec:methods_simulation}) leading to long rentals until a person returns to the starting location of their trips. However, it is worth noting that this discrepancy in distribution, while present, remains within acceptable bounds for our study. Importantly, it leads to a conservative estimation of vehicle availability for V2G applications.



In \autoref{fig:zscorestation} and \autoref{fig:validation_eventbased}, we observe that our approach generates realistic data of similar quality as an event-based simulator.
Note that it is unclear how to transfer an event-based simulator to future populations since it does not consider population statistics. Instead, one would have to vary the event rate and other parameters for simulating future scenarios. We, therefore, argue that 1) our approach yields comparable results for 2019 and 2) more feasible than an event-based approach to simulate realistic data for 2030.

\FloatBarrier
\subsection{Comparison to Inverse Reinforcement Learning}
For the sake of comparison, we perform the same analysis for the simulation with an Inverse Reinforcement Learning mode choice model. Our best-performing IRL agent shows inferior performance in predicting the \textit{correct} action  (accuracy of 42\%). It, however, also yields realistic data in distribution, although they diverge further from the real data than the data simulated with the XGB model. For brevity, we do not visualize these distributions but only provide corresponding metrics in \autoref{tab:wasserstein}. Specifically, we quantify the difference between the real and simulated distribution of duration, distance, and reservation start and end time, by means of the Wasserstein distance~\cite{vallender1974calculation}, a metric from Optimal Transport theory that is commonly used to compare probability distributions. 
Comparing IRL and XGB, we observe that the Wasserstein distances are higher for IRL, together with a mode share that diverges more from the MOBIS data, and a different distribution of reservations over stations (i.e., higher z-scores). These results indicate that the IRL model can not grasp travel behavior, or particularly the factors that affect car sharing behavior, as precisely as a supervised model. We hypothesize that the advantages of an IRL approach, namely to learn realistic state transitions over longer episodes, are negligible in our problem because the states are hardly affected by the action in our setting. Therefore, we use the supervised XGB approach for the following experiments.
\begin{table}[htb]
\resizebox{\textwidth}{!}{
\begin{tabular}{l|rrrr|r|r|r|r}
\toprule
{} &  \multicolumn{4}{c}{Wasserstein distance between real and sim} &  Avg. abs.  &  Avg. mode  &  Accuracy &  Balanced \\
\textbf{Model} &       Duration &  Driven km &  Start time &  End time                         &          station z-score              &      share ratio     &   & Accuracy             \\
\midrule
IRL   &                    1.62 &                    17.00 &                   4931.43 &                 8470.77 &                       1.21 &                   2.41 &      0.42 &           0.14 \\
XGB   &                    1.05 &                    14.62 &                   2350.14 &                 4973.33 &                       1.04 &                   1.52 &      0.78 &           0.66 \\
\bottomrule
\end{tabular}
}
\caption{Comparing the simulated data with the XGB model to the data simulated with IRL, in terms of their match with the real car sharing data. The Wasserstein distances between real and simulated distributions are higher for IRL, indicating a better fit of the XGB model. Similarly, the accuracy of predicting the correct mode on test data is better for the supervised XGB model, and mode share and the distribution over stations closer match the real data.}
\label{tab:wasserstein}
\end{table}


\FloatBarrier

\subsection{Car sharing scenarios for 2030}

\subsubsection{Scenario design}

To design realistic yet diverse scenarios for 2030, we adopt an approach by \cite{geum2014combining} who propose to combine growth scenarios with possible technological roadmaps of a company. \cite{geum2014combining} use a car sharing business as an illustrative example and simulate three scenarios for the prospective demand (pessimistic, optimistic, and neutral) and four scenarios as strategic roadmaps ("defender", "prospector", "reactor" and "analyzer"). The scenarios correspond to all possible combinations of these two parts, leading to 12 scenarios in the example. Following this approach, we assume different growth rates for the customer base (slow, intermediate, and fast growth), and design four possible  roadmaps for the company, listed in the following:
\begin{itemize}
    \item \textit{User-centered}: The number of vehicles is increased proportionally to the number of users.
    \item \textit{V2G-affine}: More vehicles than necessary are deployed, in order to enable ancillary services.
    \item \textit{Restrictive}: Less vehicles are deployed to increase the utilization rate.
    \item \textit{Expand}: New stations are installed.
\end{itemize}

Since the analysis of $12$ scenarios $(3\times 4)$ becomes convoluted, we only combine the \textit{user-centered} roadmap with the three business-growth scenarios and simulate the other business roadmaps only for the fast-growth scenario. This procedure yields the following scenarios, where we abbreviate the number of users with $U$, the number of deployed vehicles with $V$, and the number of stations with $S$:
\begin{itemize}
    \item \textbf{Scenario 1: Slow growth - \textit{User-centered} $(\times1.15)$}: 115k U, 3500 V, 1750 S
    \item \textbf{Scenario 2: Intermediate growth - \textit{User-centered} $(\times 1.5)$}: 150k U, 4500 V, 1750 S
    \item \textbf{Scenario 3: Fast growth - \textit{User-centered} $(\times 2.5)$}: 250k U, 7500 V, 1750 S
    \item \textbf{Scenario 4: Fast growth - \textit{Restrictive}}: 250k U, 5000 V, 1750 S
    \item \textbf{Scenario 5: Fast growth - \textit{V2G-affine}}: 250k U, 10000 V, 1750 S
    \item \textbf{Scenario 6: Fast growth - \textit{Expand}}: 250k U, 7500 V, 3000 S
\end{itemize}

In order to specify the exact number of users for each scenario, we considered the past growth rates of the car sharing business. Currently, there are around 3000 vehicles deployed at 1750 stations, and the validation study showed that the current user behavior matches with a scenario of around 100k users. In the past years\footnote{based on annual reports from 2017 - 2022, available at \url{https://www.mobility.ch/en/mobility-cooperative/company-reports} (last accessed: 1.9.2023)}, the car sharing service has successfully increased its customer base (8\% growth per year on average), while only slowly deploying more stations and vehicles (yearly growth rates of 0.4\% and 0.9\% respectively). However, in discussion with the company it was decided that a proportional growth of vehicle and customer number should be considered as the base scenarios for the next years. Furthermore, it is unlikely that the high customer growth rate of 8\% can be maintained over the next years. Instead, the most pessimistic scenario should assume growth of 15\% until 2030 (i.e., a yearly increase of 1.76\%), while the most optimistic scenario should assume fast growth to 2.5 times the current customers in 2030, corresponding to a yearly increase of 12.1\%.

\subsubsection{Vehicle and station utilization in future scenarios}

\begin{figure}[htb]
    \centering
    \includegraphics[width=\textwidth]{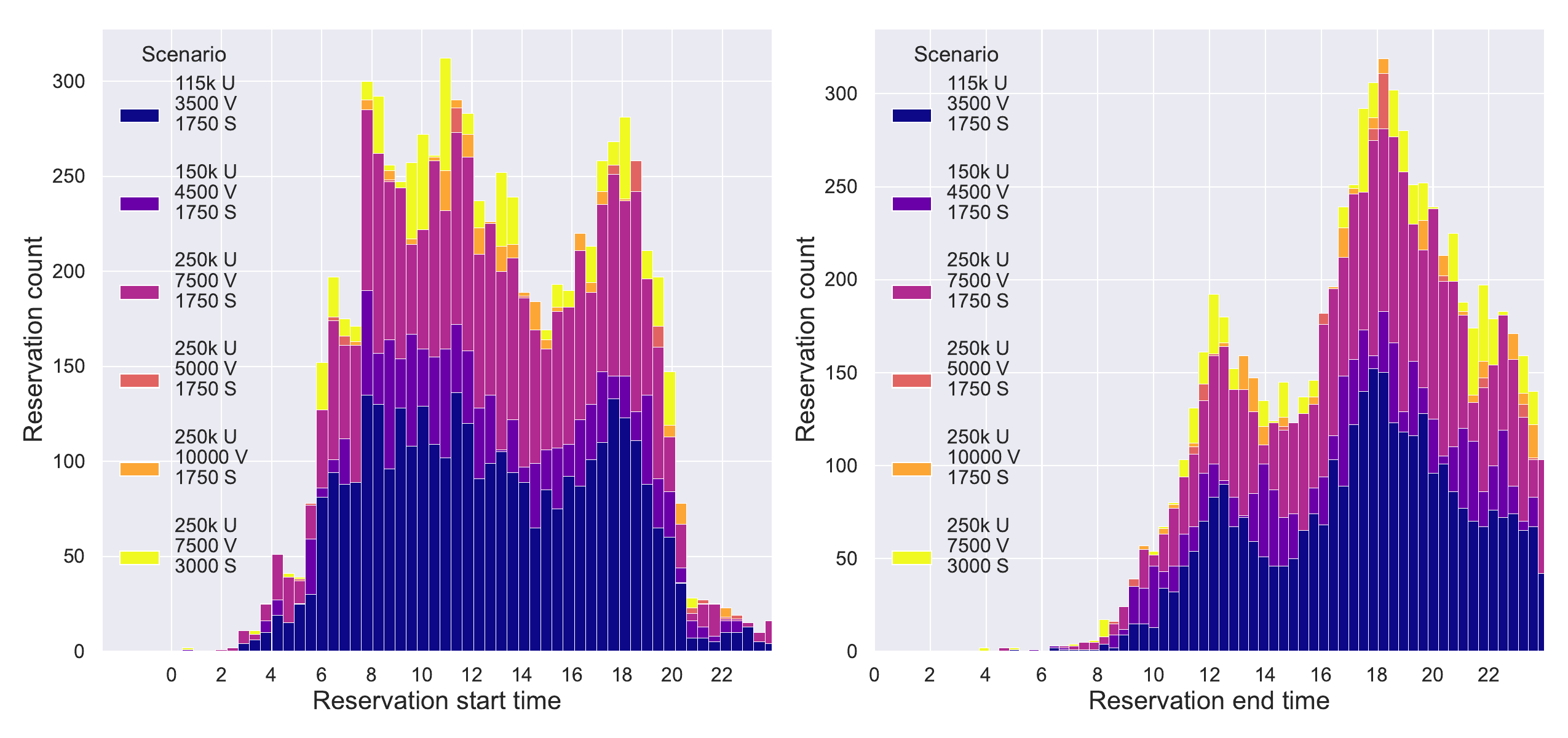}
    \caption{Distribution of reservation start and end time for all scenarios. If the number of users and vehicles increases, there are more reservations, which are distributed similarly over the day. A higher or lower number of vehicles does not have a strong effect (compare red (5000 V) and orange (10000 V) to pink (7500 V)). Additional stations (yellow) lead to more bookings at peak times.}
    \label{fig:all_scenarios}
\end{figure}

\autoref{tab:comp_2030} compares the scenarios in terms of their induced booking- and utilization rates. When scaling the number of customers and vehicles proportionally (Scenarios 1-3, \textit{User-centered}), the number of reservations also increases proportionally. \autoref{fig:all_scenarios} visualizes the number of reservations for all scenarios and contrasts them in terms of the reservation start and end time. The distributions are similar for all scenarios, but a higher number of stations (Scenario 6 \textit{Expand} - yellow) has a stronger effect than increasing the fleet size (Scenario 5 \textit{V2G-affine} - orange). We further evaluate the scenarios by their vehicle utilization rate as proposed by \cite{gonzalez_utilization_2021}. In \autoref{tab:comp_2030}, the vehicle utilization rate is given in terms of \textit{count} (what fraction of vehicles is booked at least once during that day) and \textit{time} (how many hours those vehicles are booked on average). 
When the number of users and vehicles increases simultaneously (Scenarios 1-3, \textit{User-centered}), the utilization rate is constant (61\% - 64\% of the vehicles are used around 34\% of the time). Other attributes of the reservations, such as the start and end time, also remain similar in distribution (see \autoref{fig:all_scenarios}).

\begin{table}[ht]
    \centering
\resizebox{\textwidth}{!}{
\begin{tabular}{llllccccc}
\toprule
      &     &      &  
    &  Number of & \multicolumn{2}{c}{Vehicle utilization} & Station utilization & Average distance to \\
&        &     &      &  
        reservations &  (time) &  (count) & (count) &  station (trip origin) \\
Scenario & Users & Vehicles & Stations &                         &                             &                              &                              &                              \\
\midrule
1 & 115 & 3500  & 1750 &                    3437 &                        0.34 &                         0.64 &                         0.85 &                       741.23 \\
2 & 150 & 4500  & 1750 &                    4450 &                        0.34 &                         0.61 &                         0.88 &                       748.31 \\
3 &  250   & 7500  & 1750 &                    7156 &                        0.35 &                         0.58 &                         0.93 &                       782.47 \\
4 & 250 & 5000  & 1750 &                    6933 &                        0.39 &                         0.75 &                         0.96 &                       738.65 \\

5 &  250  & 10000 & 1750 &                    7300 &                        0.33 &                         0.47 &                         0.92 &                       801.35 \\
6 &  250  &  7500     & 3000 &                    7763 &                        0.34 &                         0.65 &                         0.85 &                       512.24 \\
\bottomrule
\end{tabular}
}

    \caption{Comparing scenarios for 2030 in terms of reservation count and vehicle utilization}
    \label{tab:comp_2030}
\end{table}


\begin{figure}[htb]
    \centering
    \begin{subfigure}[b]{0.56\textwidth}
    \includegraphics[width=\textwidth]{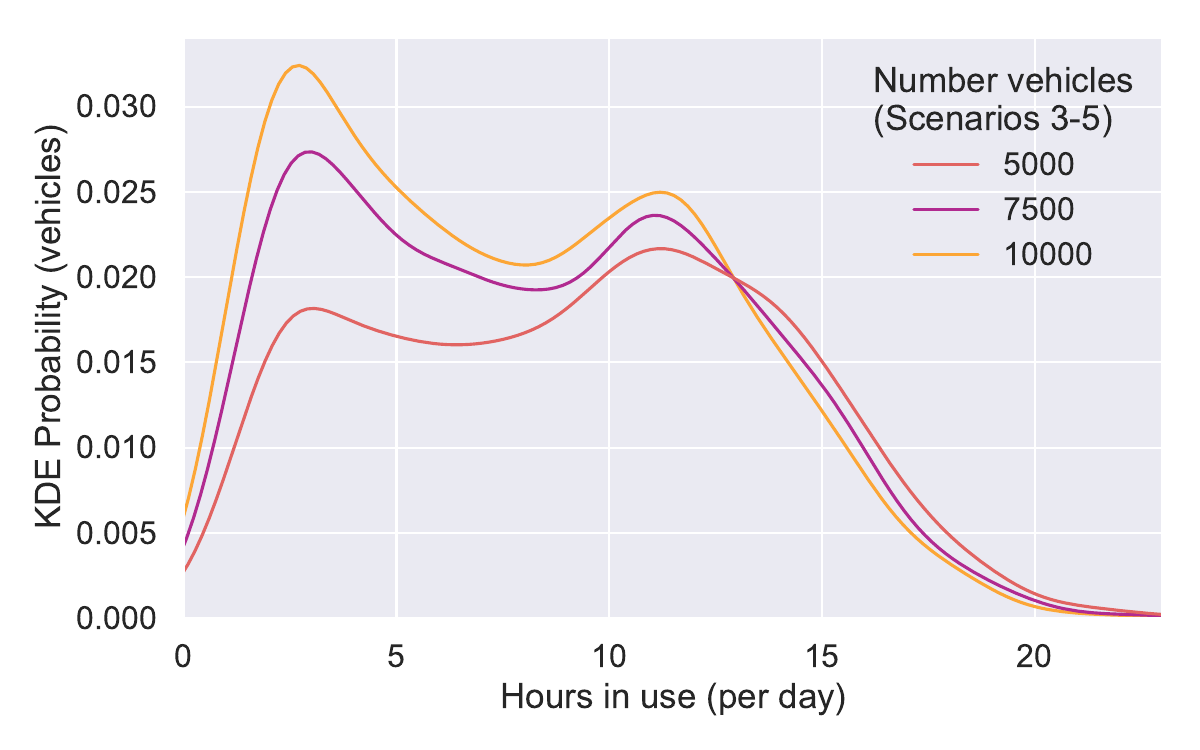}
    \caption{Distribution of duration}
    \label{fig:vehicle_util_time}
    \end{subfigure}
    \hfill
    \begin{subfigure}[b]{0.42\textwidth}
    \includegraphics[width=\textwidth]{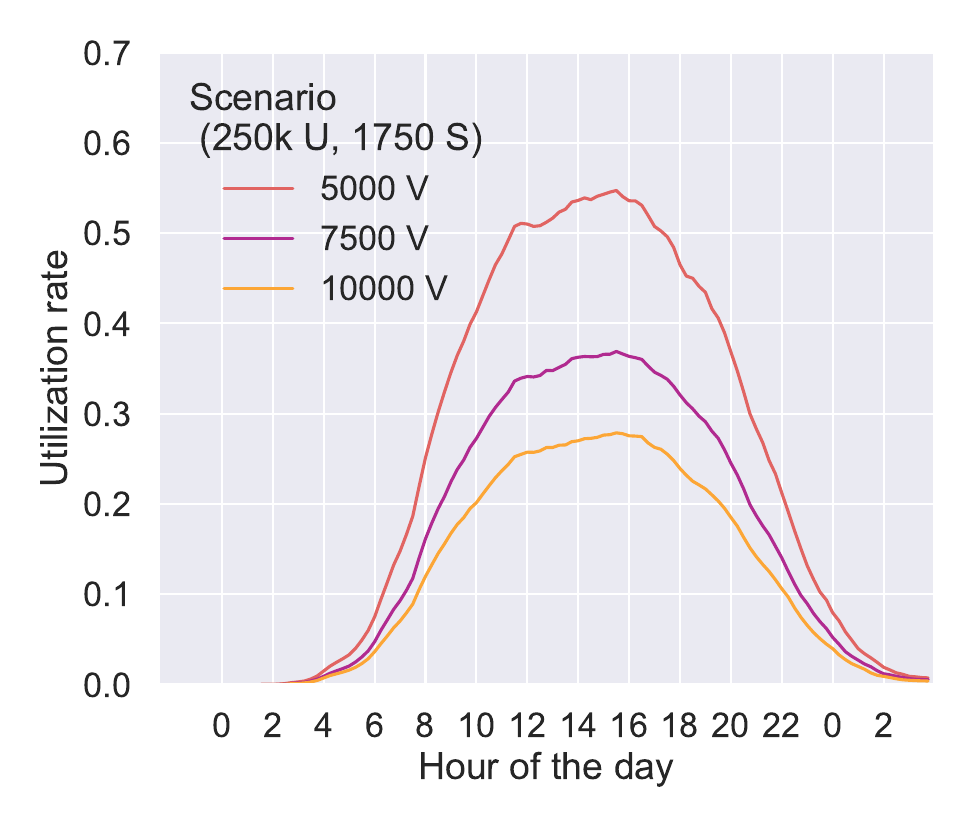}
    \caption{Vehicle utilization}
    \label{fig:vehicle_util_count}
    \end{subfigure}
    \caption{Comparing utilization rates by the number of deployed vehicles. All three scenarios assume 250k active users and 1750 stations. As expected, the utilization rate decreases if more vehicles are deployed.}
    \label{fig:scenario_vehicle_comp}
\end{figure}

Interestingly, reducing or increasing the number of vehicles (Scenario 4 \textit{Restrictive} and Scenario 5 \textit{V2G-affine}) only has a minor effect on the number of reservations in one day, decreasing and increasing it by 3.12\% and 2.0\% respectively in comparison to Scenario 3 (\textit{User-centered}).
As a result, the vehicle utilization rate changes in these scenarios, as shown in \autoref{fig:scenario_vehicle_comp}. While the shape of the distribution over a day is similar, Scenario 4 with only 5000 deployed vehicles leads to utilization rates of more than 50\% (see \autoref{fig:vehicle_util_count}). In other words, a limited availability of vehicles makes users choose other vehicles (possibly at other stations) that are still available. The main effect of Scenario 4 is an increase of the \textit{number} of vehicles that are used on the simulated day (75\% compared to 58\%, see \autoref{tab:comp_2030}), together with an increase of the \textit{duration} they are used on average (\autoref{fig:vehicle_util_time}).  
However, the simulation does not consider the time necessary for refueling or recharging the car in between its usage. High utilization of vehicles, as in Scenario 4, might therefore be infeasible in practice. Thus, we hypothesize that in a real environment, reducing the number of deployed vehicles can result in lower reservation numbers than in our simulation.

\begin{figure}[htb]
    \centering
    \begin{subfigure}[b]{0.42\textwidth}
    \includegraphics[width=\textwidth]{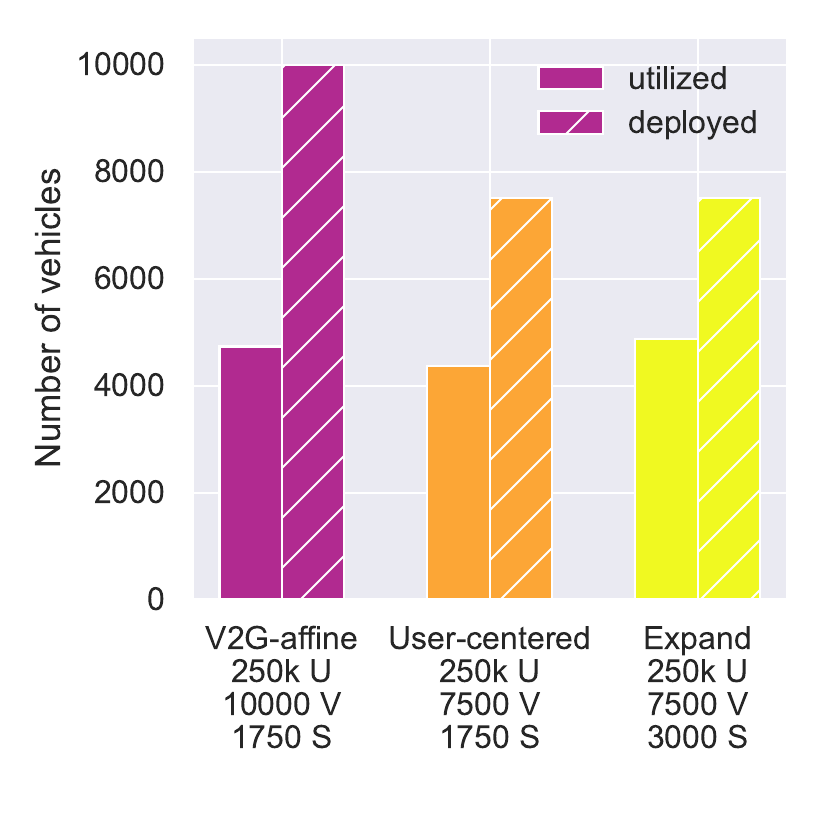}
    \caption{Vehicle utilization}
    \label{fig:station_scenario_comp}
    \end{subfigure}
    \hfill
    \begin{subfigure}[b]{0.54\textwidth}
    \includegraphics[width=\textwidth]{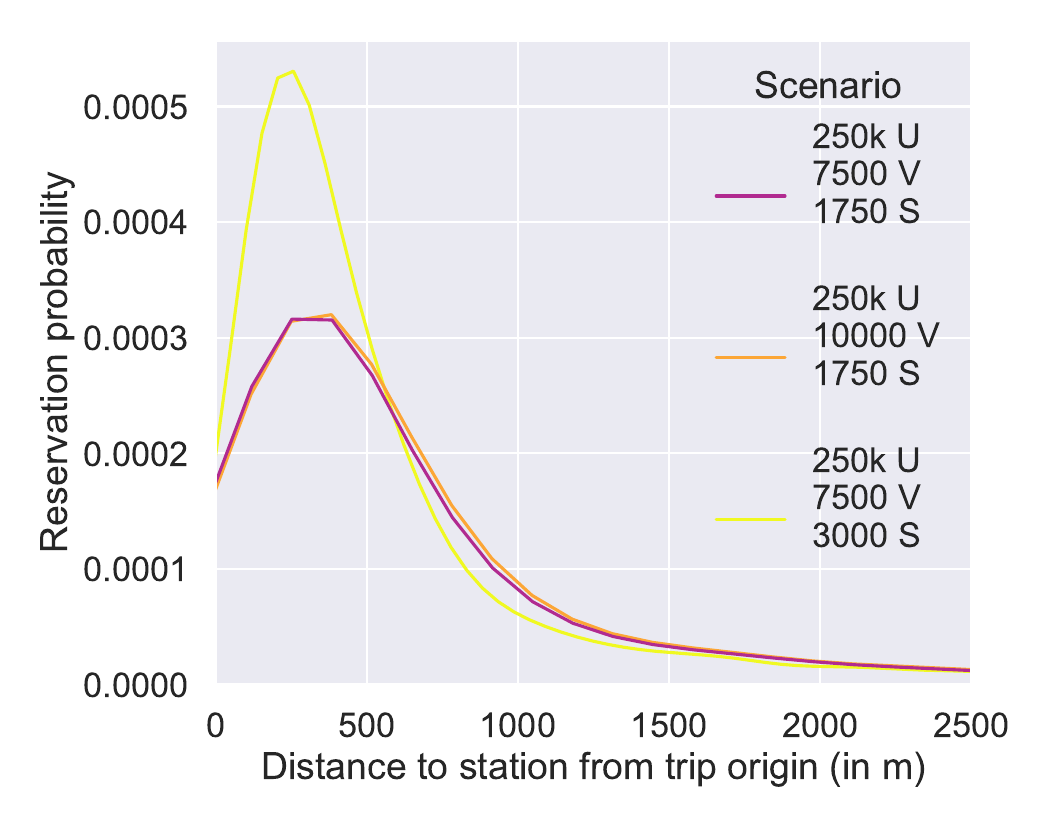}
    \caption{Distance to station}
    \label{fig:distance_station}
    \end{subfigure}
    \caption{Comparing scenarios with 1750 stations and varying vehicle numbers to a scenario with 3000 stations. The new stations cause many new reservations, since the distance of the user to the closest station decreases (\subref{fig:distance_station}). The effect of new stations is larger than the one of more available vehicles (\subref{fig:station_scenario_comp}).}
    \label{fig:scenario_station_comp}
\end{figure}

\autoref{fig:scenario_station_comp} analyzes the effect of new stations instead of additional vehicles by comparing Scenario 3 (\textit{User-centered}, orange), 5 (\textit{V2G-affine}, pink) and 6 (\textit{Expand}, yellow). In contrast to deploying more vehicles (Scenario 5), adding new stations (Scenario 6) causes significantly more bookings, namely 7763 instead of 7156 ($+8.5$\%), as shown in \autoref{fig:station_scenario_comp}. 
The reason is the dependency of the user's mode choice on the distance of the closest available vehicle. When placing new stations, the activities of the synthetic population are located closer to a car sharing station. This is quantified in the column ''Average distance to station (trip origin)'' in \autoref{tab:comp_2030}. In Scenario 6 (3000 S), the location of the trip origin is 512 meters away from the closest station on average, whereas it is more than 730 meters in other scenarios (1750 S). The corresponding distribution is shown in \autoref{fig:distance_station}. 
Despite the large number of stations, the station utilization of Scenario 6 is still high, with 85\% of the stations being visited at least once during the simulated day (see \autoref{tab:comp_2030}). The vehicle utilization of Scenario 6 is only surpassed by Scenario 4 (\textit{Restrictive}). 

\FloatBarrier

\section{Future opportunities for vehicle-to-grid in car sharing fleet}

It was demonstrated that the roadmap of the car sharing business, together with its general growth rate, will strongly impact the utilization rates and the spatio-temporal distribution of mobility demand. Taking this one step further, we aim to analyze the resulting potential for ancillary services. We consider the same scenarios for 2030 and schedule V2G operations to understand its maximum potential under the constraints of user reservations. As explained above, we use the optimization framework by~\cite{nespoli2022national} to schedule charging and discharging operations over one day. The objective function in this framework is flexible and can be designed to minimize costs and charging times, to make maximum use of an associated photovoltaic system, or to achieve peak shaving for the (regional) Distribution System Operator (DSO). Here, we consider the business case where the fleet provides ancillary services to a DSO. The services are paid based on a preceding bidding process; i.e., the fleet owner bids the capacity of the fleet on the electricity market and the DSO agrees on a price per megawatt (MW) to compensate for the power flexibility supplied by the fleet. 
\begin{wrapfigure}[19]{r}{0.5\textwidth}
  \begin{center}
  \vspace{-1em}
    \includegraphics[width=0.48\textwidth]{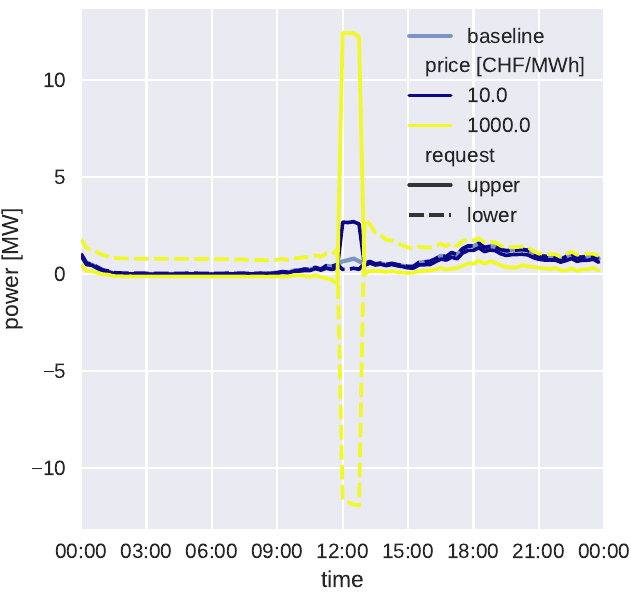}
  \end{center}
  \caption{Flexibility}
  \label{fig:flexi_example}
\end{wrapfigure}
Note that the optimization of the prices by itself is a separate research field~\cite{sortomme2011optimal, ansari2014coordinated}; here, we simply test the effect for a range of prices. We first quantify how much flexibility can be provided maximally throughout the day, which can help specify the available power in the bidding process. Secondly, we conduct a case study in a single DSO to quantify the peak shaving effect.

\subsection{Potential flexibility of the car sharing fleet}

We follow \cite{nespoli2022national} and \cite{oldewurtel_towards_2013} to implement an experiment that aims to answer the following question: 
How much power is available for ancillary services, at what time and at which price? For this purpose, the optimization problem is solved for every hour of the day with respect to a fleet-level objective that minimizes the distance to a reference profile of the power consumption~\cite{oldewurtel_towards_2013}. \autoref{fig:flexi_example} provides an example of the response of the fleet in Scenario 1 (150k users, 3500 vehicles, 1750 stations) for a single hour. The higher the price paid to the fleet operator, the more power is provided. With a very large price of 1000 CHF/MW, up to 12 MW can be provided by the car sharing fleet. Note that the optimization algorithm guarantees that all user reservations remain feasible in terms of sufficient State-of-Charge (SOC) and vehicle availability. 

\begin{figure}[ht]
    \centering
    \includegraphics[width=\textwidth]{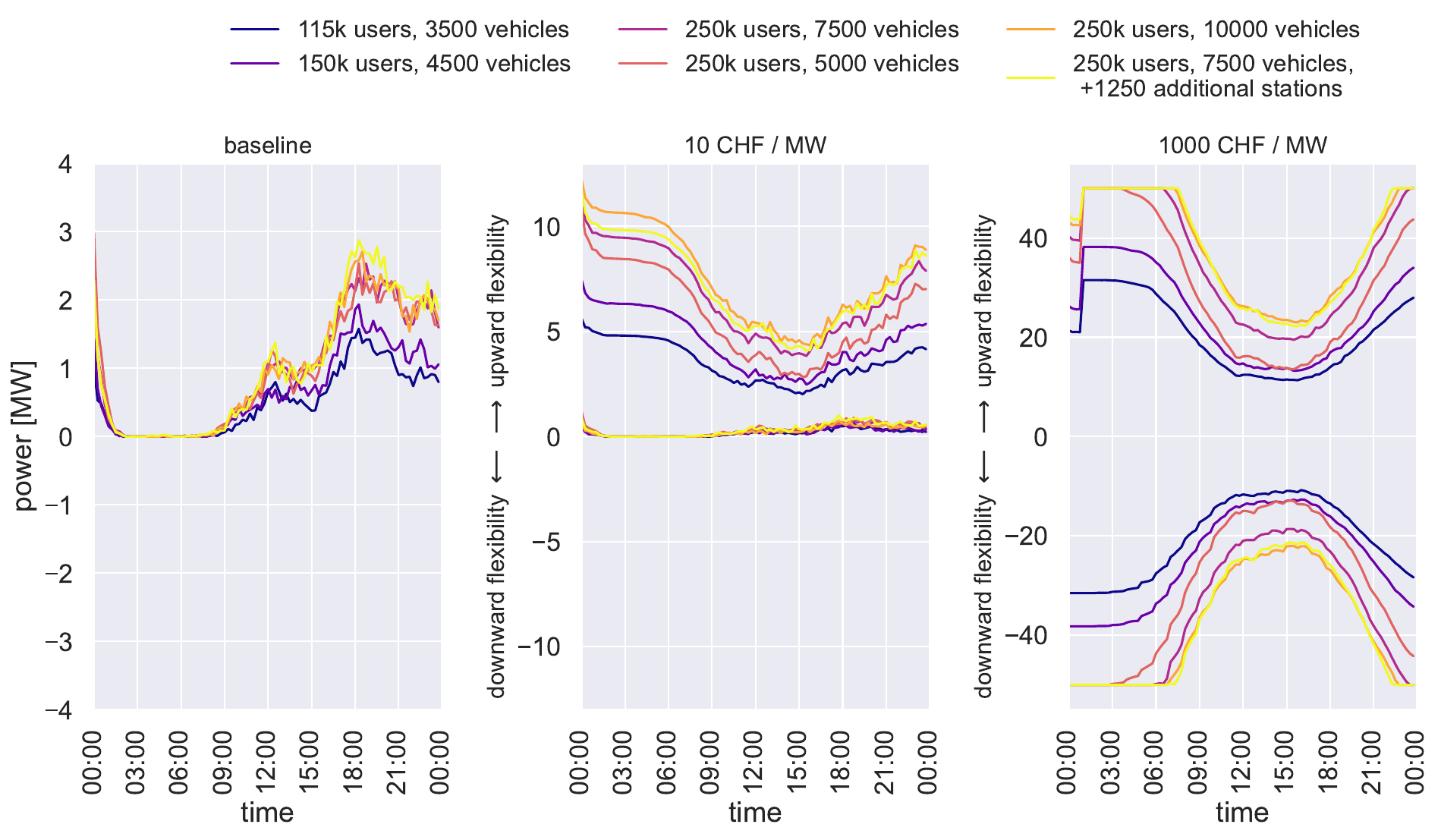}
    \caption{Up- and downward flexibility of the fleet in different price levels and car sharing scenarios.
    }
    \label{fig:up_down_flexibility}
\end{figure}

Computing the flexibility response for each hour of the day yields an envelope that bounds the upward and downward flexibility as shown in \autoref{fig:up_down_flexibility}. The baseline scenario (no payment for ancillary services) simply reflects the charging behavior of the fully-electric car sharing fleet. The charging power peaks in the evening at 6pm. With a low price of 10 CHF/MW, up to 10 MW charging flexibility is feasible, depending on the scenario; however, V2G is not profitable for the fleet owner at that price. The lowest flexibility is achieved in the afternoon due to the peak usage of cars (compare \autoref{fig:vehicle_util_count}). With 1000 CHF/MW, there is a strong V2G response of up to 50 MW discharging. In general, the more vehicles in the system, the higher the potential for charging and discharging control. It is further shown that scenarios with a lower number of vehicles, especially Scenario 4 (\textit{Restrictive}, 5000 vehicles), result in less flexibility, in particular at peak times. Finally, scenario 6 with 1250 additional stations achieves almost the same flexibility as scenario 5 with 10000 vehicles, despite the induced demand that leads to a higher number of car sharing reservations in this scenario (see \autoref{tab:comp_2030}. The reason is the higher number of charging stations, which simplifies V2G scheduling. 



\FloatBarrier

\subsection{Peak-shaving potential}

Secondly, we analyze the peak-shaving potential of V2G in future car sharing scenarios. Since the energy demand data are not publicly available for every region in Switzerland, we restrict this experiment to stations within the Distribution System Operator (DSO) of Zürich, named Elektrizitätswerk der Stadt Zürich (EWZ), which is the DSO that covers the most car sharing stations (246 out of 1750 car sharing stations, with 726 vehicles in 2019) and that provides public data~\cite{ewz_bruttolastgang_stadt_zuerich}. To incentivize peak shaving, the fleet-level objective is designed to punish any increase in the DSO's peak load and to reward a reduction of the peak load in the considered period. Note that there is no reward if the current energy demand is lower than at an earlier point in the time period. The results must be taken with a grain of salt since the one-day simulation is not representative of the realistic scenario of peak-shaving on a monthly basis.

\begin{figure}[ht]
    \centering
    \includegraphics[width=\textwidth]{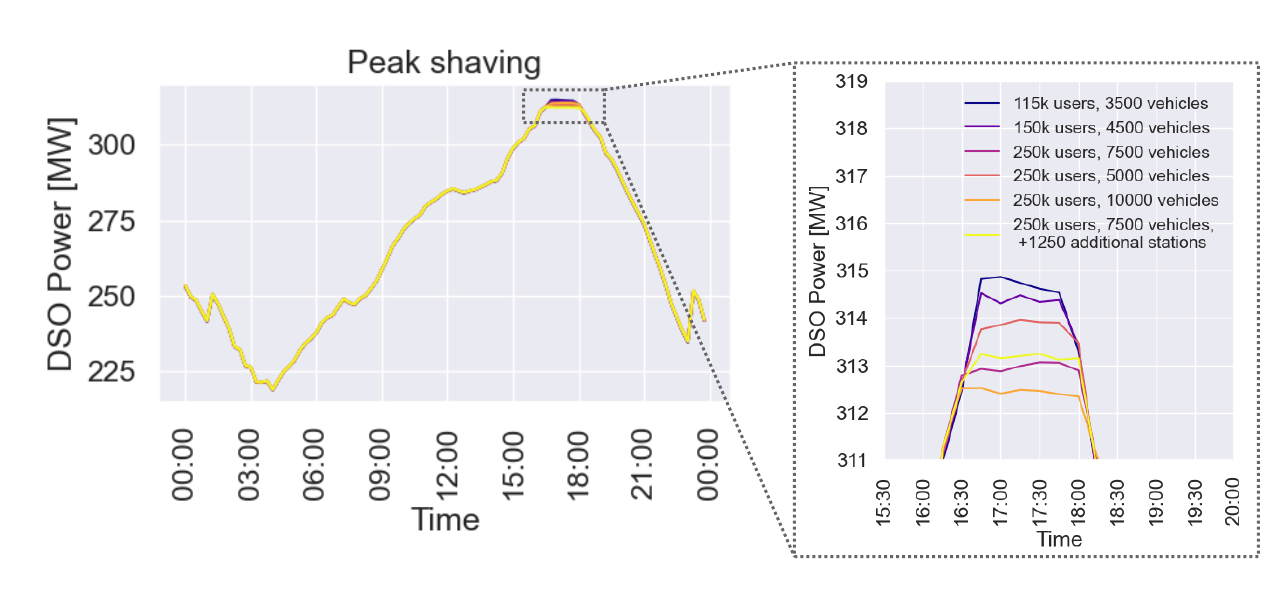}
    \caption{Peak shaving case study by the example of the DSO of Zurich. The power flexibility at a price level of 1000 CHF/MW is shown for all scenarios.
    }
    \label{fig:peak_shaving}
\end{figure}

\autoref{fig:peak_shaving} visualizes the peak shaving on a simulated day. While the effect appears small with respect to the overall power consumed in the DSO, the high cost of peak loads justifies the business case. The peak can be shaved by up to 4.6 MW in scenario 5 (10000 vehicles), where the baseline peak load of 317.12 MW is reduced to 312.52 MW. This is remarkable in a single DSO that covers only a minor part of the car sharing fleet. Meanwhile, there is a smaller peak shaving effect in Scenario 6 (1250 additional stations), contradicting the previously reported results testifying a high flexibility for this scenario. The reason for this mismatch is the higher demand for vehicles in the urbanized areas within the DSO in Zürich, leading to lower energy flexibility at the peak time in the evening.

\begin{figure}[ht]
    \centering
    \includegraphics[width=\textwidth]{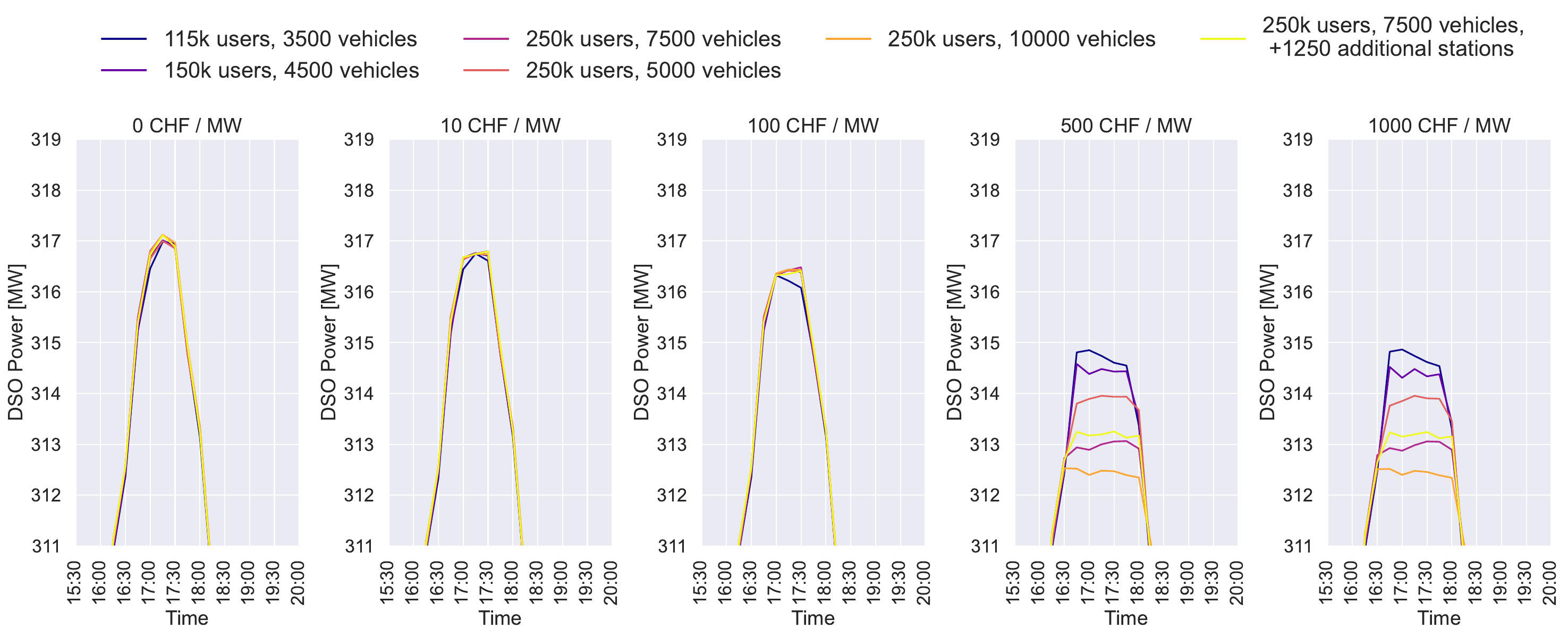}
    \caption{Peak shaving for one DSO, by the price per MW.}
    \label{fig:peak_shaving_price}
\end{figure}

Furthermore, \autoref{fig:peak_shaving_price} distinguishes the peak-shaving effect by the price level. Only minor effects ($<1$ MW) are achieved when the price is less or equal to 100 CHF/MW. The peak is even visibly reduced simply by a lower number of deployed electric vehicles in Scenario 1 (blue line), in contrast to actual peak shaving. With 500 or more CHF/MW, there is a strong response that converges to a peak load with as low as 312.5 WM. 

\FloatBarrier

\begin{figure}
    \centering
    \begin{subfigure}[b]{0.48\textwidth}
    \includegraphics[width=\textwidth]{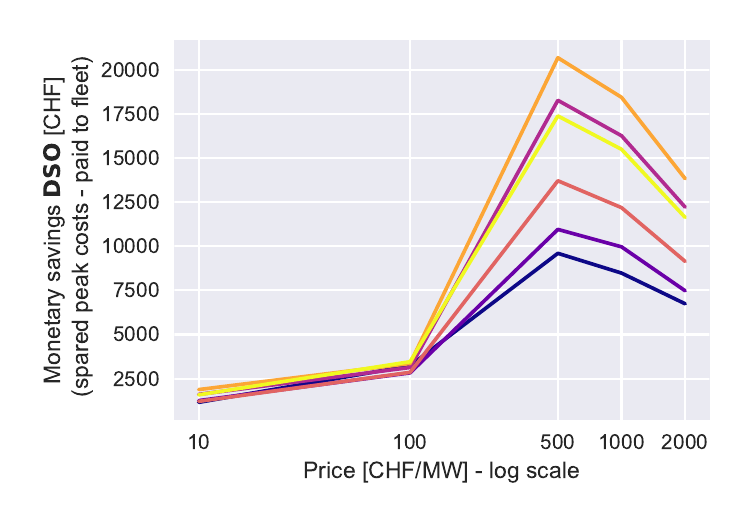}
    \end{subfigure}
    \begin{subfigure}[b]{0.48\textwidth}
    \includegraphics[width=\textwidth]{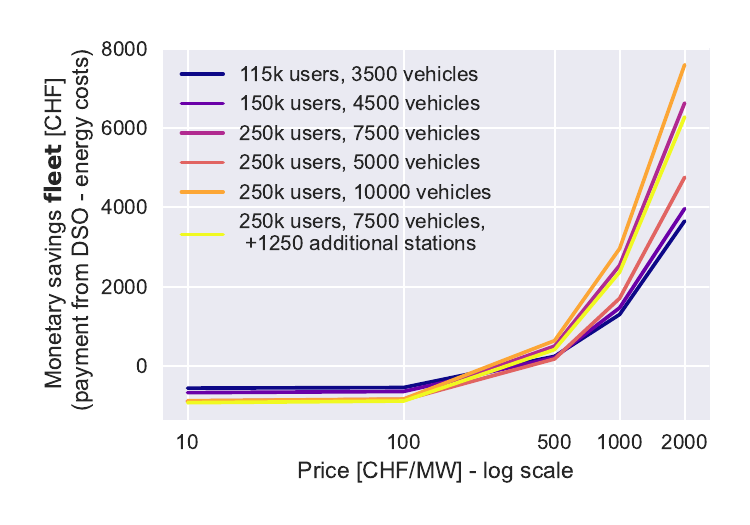}
    \end{subfigure}
    \caption{Monetary savings for DSO and fleet owner involved in V2G.}
    \label{fig:monetary}
\end{figure}

Finally, we quantify the gains from peak shaving in terms of monetary savings for the car sharing fleet and the DSO in \autoref{fig:monetary}. The DSO benefits from the lower peak energy demand but must pay the fleet for the ancillary services. The fleet earns from the compensation, but has increased energy costs, if the grid tariffs are not reimbursed for V2G services\footnote{Reimbursement for grid tariffs is not in place in Switzerland, but exists in other countries, for example, Italy~\cite{ARERA_2022}, and might therefore be available in 2030.}. This trade-off is shown in \autoref{fig:monetary}. The DSO benefits most when the agreed price is 500 CHF/MW, where up to 20k CHF is saved in the scenario with 10000 vehicles. On the other hand, the fleet earns more with increasing compensation, and, considering the energy costs, only profits for prices of more than 500 CHF / MW. 
Together, a sweet spot is between 500 and 2000 CHF where both fleet and DSO clearly benefit from implementing V2G for the car sharing fleet.

\section{Discussion}\label{sec:discussion}

V2G has long been discussed as a promising technology in the transition to sustainable energy and transportation, but its potential for shared vehicles has been neglected so far. To the best of our knowledge, our scenario-based study is the first attempt to quantitatively assess the future potential of V2G in car sharing. Overall, we have found strong evidence supporting a business case of V2G for large car sharing fleets. Even in a slow-growth scenario (with 115k users, 3500 vehicles, 1750 stations), the car sharing service can provide more than 10 MW flexibility for ancillary services, even at times of high vehicle demand. More optimistic scenarios for the car sharing service lead to a V2G potential of up to 50 MW on a national scale, which marks a significant contribution to the power system stability. A more in-depth analysis of a local DSO, covering 246 car sharing stations, revealed a peak-shaving potential with benefits for both the car sharing fleet and DSO. In all scenarios, there is a sweet spot where both DSO and the car sharing operator benefit from the trade. In our case study, the sweet spot is between prices of 500 CHF/MW and 2000 CHF/MW; however, the optimal price level depends strongly on the peak load costs of the DSO, as well as the involved electricity costs for the car sharing fleet, that might be lowered due to regulatory measures in the future.

Our methodological contribution is a data-driven pipeline that simulates car sharing with consideration of the complex interplay between population socio-demographics, user behavior, system changes, and business decisions. In contrast to event-based (sampling-based) simulations, our agent-based framework reflects changes in the population and spatial factors, such as the proximity of agents to a car sharing station. In contrast to other agent-based simulators, we implemented a data-driven approach that captures the complexity of human decision-making with state-of-the-art ML techniques, considering a wealth of features affecting transport mode choices, such as the availability of other transport modes, daytime, and trip purpose. It remains worthwhile for follow-up work to use more generic traffic simulators such as MATSim to account for the interplay between different transport modes, such as congestion. It is worth noting that this probably comes at the cost of requiring to subsample of the population. The results of our car sharing simulation revealed interesting patterns in the utilization of car sharing vehicles. 
In particular, the number of deployed vehicles mainly affected the utilization rate, instead of the number of reservations. Our simulation further shows a positive side effect of deploying additional stations, inducing an 8.5\% increase in the reservation count. Therefore, the installation of new stations may be a promising endeavor for the car sharing operator, leading to higher car sharing demand coupled with more flexibility for V2G due to additional charging stations.

The scenarios analyzed aim to encompass a diverse range of growth scenarios, business changes, and electricity price levels. However, the scenarios are not extensive and cannot account for shifts in travel \textit{behavior} unrelated to demographic or infrastructure-induced changes. Most importantly, the usage patterns of \textit{electric} car sharing may differ from those observed for combustion engine vehicles in our dataset, where only 3.5\% of trips involved EVs. Therefore, replicating our study using an electric car sharing dataset could be a valuable next step. Furthermore, the MOBIS dataset, utilized in this study for mode choice modelling, was collected before and during the COVID-19 pandemic. Our experiments on test data from 2022 verified the model's applicability for predicting post-COVID mobility behaviour, in accordance with other studies that report only small changes in Switzerland's modal split due to the pandemic~\cite{molloy2021observed, heimgartner2023modal} (see \autoref{tab:pre_post_covid} in Appendix~\ref{app:mode_choice_stability}). Considering that commuting accounts for merely 1\% of station-based car sharing trips~\cite{becker2017comparing}, the pandemic-induced increase in remote work has minimal impact on car sharing patterns. These observations support the applicability of the mode choice model to scenarios in 2030; yet future research should aim to capture attitude changes or trends that alter car sharing usage or adoption beyond an increasing number of users; e.g., car sharing attracting new target groups. 

Furthermore, 
a key question for future research is the impact of V2G operations on spontaneous bookings. The scheduling algorithm employed in this work assumes all bookings to be known with the goal of quantifying the \textit{maximum} potential for V2G. However, in the Swiss car sharing system, around 20\% of the bookings are made less than half an hour before the reservation time. Ancillary services might in this case prevent the reservation and reduce customer satisfaction when the SOC is insufficient for the desired trips. While ensuring the transparency of the SOC via digital platforms is one promising solution, there still exists the additional challenge for the car sharing service to dynamically reschedule V2G services if a specific vehicle is suddenly desired by a customer. Thus, an optimal planning strategy, like Model Predictive Control~\cite{garcia1989model}, is required, where control algorithms also need to scale to a national-sized fleet.

While our results do not directly generalize to other types of car sharing systems (e.g., one-way car sharing) or other countries, they provide a justification for the potential business opportunity of V2G in car sharing. More broadly speaking, our study implies a significant contribution of car sharing for sustainable transportation, since it not only reduces car ownership, but can also support power grid stability and reduce the challenges involved in the upcoming large-scale electrification of transportation. Considering the central management of the car sharing fleet, the feasibility of implementing V2G for the car sharing fleet is higher than for private vehicles. This calls for policies regulating the electricity market with improved conditions for fleet owners, especially for car sharing fleets, considering their additional contribution to sustainable transport.

\section{Conclusion}\label{sec:conclusion}

The transition towards a sustainable economy is hardly feasible without new technologies in the energy and transportation sectors. Our study demonstrates that V2G is not only promising for decarbonizing individual transport, but also integrates well with shared services; a business model that might become more prevalent with the adoption of autonomous driving~\cite{liu2022effects}. This finding opens many research opportunities, including further scenario studies, developing scalable scheduling algorithms, leveraging predictive models for estimating transport and power demand, or improving the bidding process with digital platforms.

\section*{Acknowledgements}

We thank \textit{Mobility} for providing the dataset and feedback on the paper. 
This work was funded by the Swiss Federal Office of Energy (V2G4CarSharing project) with grant SI/502344-01. We also would like to thank the Institute for Transport Planning and Systems at ETH Zürich for their technical support in utilizing their open-source code for generating synthetic populations.

\section*{Data and Software Availability}

The source code for car sharing simulation is available at \url{https://github.com/mie-lab/car_sharing_simulator}. Further source code covering all analysis steps is provided at \url{https://github.com/mie-lab/v2g4carsharing}. The car sharing dataset cannot be shared.

\section*{Author contributions}

Nina Wiedemann: Conceptualisation, Methodology, Software, Validation, Formal Analysis, Investigation, Data Curation, Writing - Original Draft, Visualization\\
Yanan Xin: Conceptualisation, Methodology, Writing - Review \& Editing, Supervision, Project administration, Funding acquisition\\
Vasco Medici: Conceptualisation, Software, Investigation, Visualization, Writing - Review \& Editing\\
Lorenzo Nespoli: Conceptualisation, Software, Investigation, Visualization, Writing - Review \& Editing\\
Esra Suel: Writing - Review \& Editing\\
Martin Raubal: Conceptualisation, Resources, Supervision, Project administration, Funding acquisition, Writing - Review \& Editing



\bibliographystyle{elsarticle-num} 
\bibliography{refs}

\appendix
\newpage
\section*{Appendix}

    
    

\section{Comparing to an event-based car sharing simulation}\label{app:eventbased}

To implement an event-based simulation, we follow the work by \cite{cocca_free_2019, cocca_car-sharing_2020, fassio_environmental_2021}, and simulate booking-events from a Poisson process. In their work, the spatial distribution of bookings is approximated with Kernel Density Estimation, since they work on a free-floating car sharing dataset. For our purposes, a categorical distribution over the stations is sufficient. Last, the duration and distance are approximated best with an exponential power distribution of the form $p(x; \lambda, k) = x^k \cdot \exp{(-\lambda x)}$. All distributions are fitted on the real data separately for every hour of the day and distinguishing weekdays from weekend, to reflect temporal differences in the booking behaviour. This process yields 48 Poisson distributions, 48 categorical distributions over the stations and two times 48 power exponential distributions that comprehensively reflect the booking behaviour at each hour of the day.

\begin{figure}[bth]
    \centering
    \includegraphics[width=0.4\textwidth]{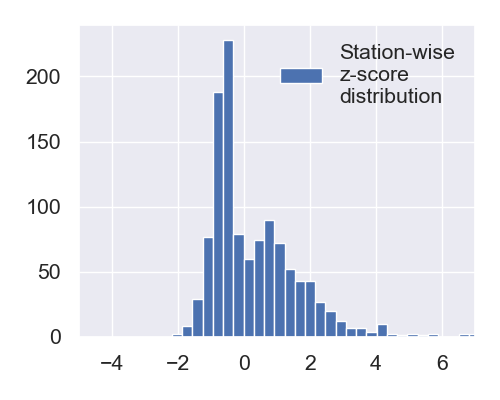}
    \caption{Station-wise z-score of the event-based simulator}
    \label{fig:event_stations}
\end{figure}

The bookings for one day are simulated by drawing the number of bookings $n(t)$ from the correct Poisson distribution at every half an hour, and then drawing $n(t)$ durations, distances and stations from the respective distribution. Note that the vehicle ID are not simulated, and it must be assumed that sufficient vehicles are available per station. In addition to \autoref{fig:validation_eventbased}, we provide the z-scores of the station-wise bookings here in \autoref{fig:event_stations}. The distribution is very similar to the one of our simulator (\autoref{fig:stationdist}), confirming that our approach yields similarly realistic behaviour as an event-based simulation.

\section{Validating distribution of simulated car sharing user population}\label{app:carsharinguser_sampling}

\begin{figure}[htb]
    \centering
    \includegraphics[width=\textwidth]{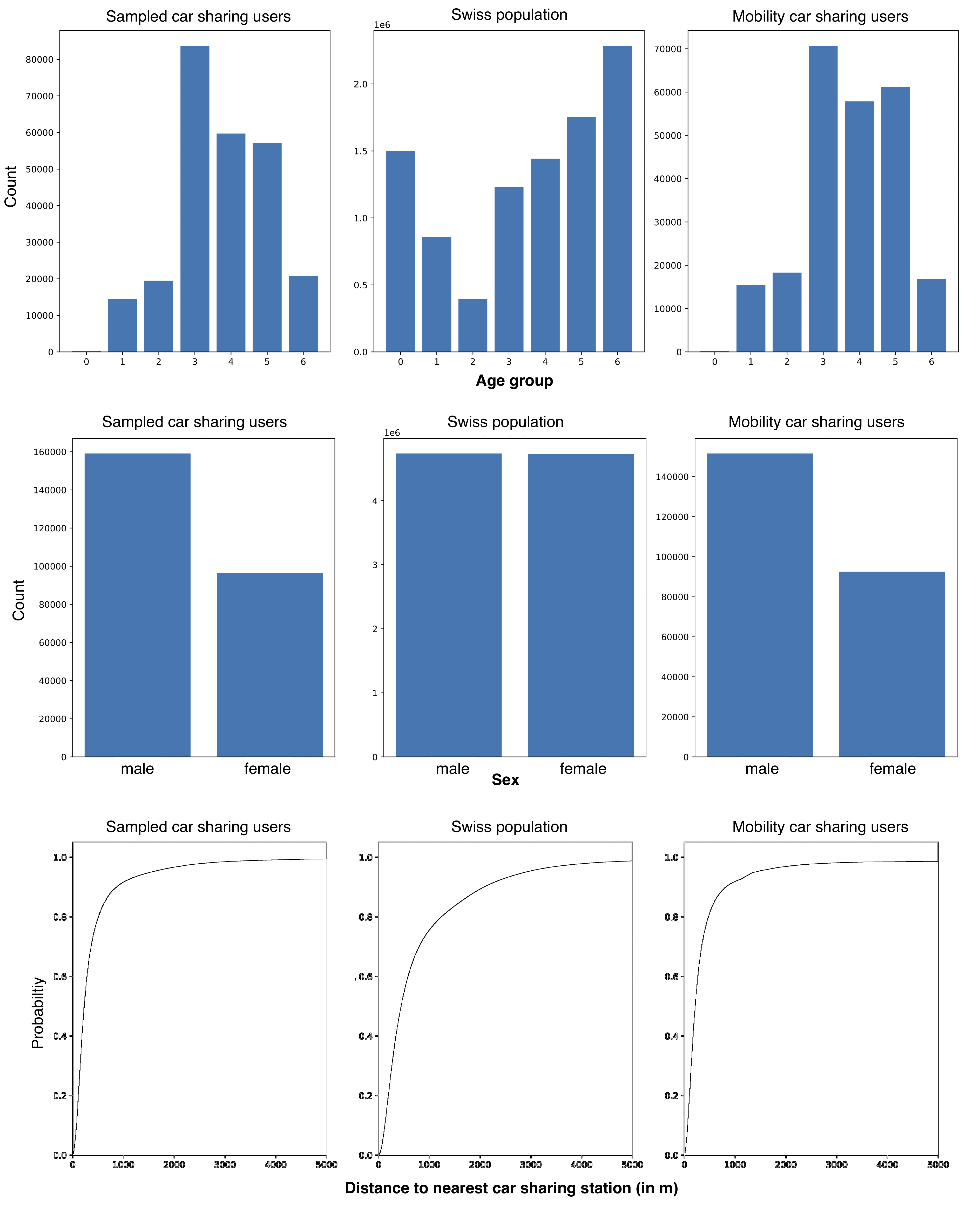}
    \caption{Comparison of general population, simulated and real car sharing users. The simulated car sharing users match the real ones in distribution}
    \label{fig:stratified_sampling}
\end{figure}

We validate our stratified sampling approach by providing the distribution of age, gender and the distance to the closest car sharing station among the simulated users in \autoref{fig:stratified_sampling}.

\section{Applicability of the mode choice model for future populations}\label{app:mode_choice_stability}

Since we train the mode choice model on data from 2019, but apply it for a simulated population in 2030, we need to validate its stability over time. For this purpose, we take the latest available excerpt of the MOBIS dataset~\cite{molloy2022mobis}, namely from June 2022 to December 2022, called 2022-data in the following. The data was collected from participants in the original MOBIS study who agreed to collect tracking data via the installed app beyond the study period. We test the mode choice model on 491 users and 153301 trips available in the 2022-data, where we selected only users that do not appear in our training data. Nevertheless, the accuracy of the model in classifying the transport mode is still 68\%, compared to 79\% test accuracy on a split of the original data. On further investigation, we noted that seasonal effects lead to significantly more bike rides during this period. It must be noted, therefore, that our model is better applicable in winter periods. Furthermore, difficulties in the prediction may also stem from different mobility behaviour of the new users compared to car sharing users (the model was fitted only on trips of car sharing subscribers). When filtering the 2022-data for car sharing users, the accuracy increases to 75\%. Since we aim to capture typical behavior of car sharing subscribers, this bias is intended. 

\autoref{tab:pre_post_covid} further shows that the COVID-19 had only minor influence on travel patterns of car sharing users in the MOBIS dataset. Apart from a marginally lower travel distance during the pandemic, the trip purpose, time, and the PT or car sharing accessibility hardly changed.

\begin{table}[htb]
    \centering
\resizebox{0.8\textwidth}{!}{
\begin{tabular}{llll}
\toprule
& \multicolumn{3}{c}{Trips of MOBIS car sharing users }\\
 & pre-COVID & during COVID & post-COVID \\
Feature &  &  &  \\
\midrule
distance & 12004.61 (19207.51) & 11851.52 (19751.06) & 12147.27 (20781.36) \\
purpose destination home & 0.36 (0.48) & 0.38 (0.48) & 0.37 (0.48) \\
purpose destination leisure & 0.25 (0.43) & 0.23 (0.42) & 0.26 (0.44) \\
purpose destination work & 0.24 (0.43) & 0.22 (0.41) & 0.18 (0.39) \\
purpose destination shopping & 0.07 (0.25) & 0.09 (0.29) & 0.07 (0.25) \\
purpose destination education & 0.02 (0.15) & 0.0 (0.07) & 0.0 (0.07) \\
purpose origin home & 0.35 (0.48) & 0.37 (0.48) & 0.37 (0.48) \\
purpose origin leisure & 0.25 (0.43) & 0.23 (0.42) & 0.26 (0.44) \\
purpose origin work & 0.24 (0.43) & 0.22 (0.41) & 0.19 (0.39) \\
purpose origin shopping & 0.07 (0.25) & 0.09 (0.29) & 0.07 (0.25) \\
purpose origin education & 0.02 (0.15) & 0.0 (0.07) & 0.0 (0.07) \\
PT accessibility (origin) & 2.13 (1.47) & 2.12 (1.44) & 2.54 (1.46) \\
PT accessibility (destination) & 2.12 (1.47) & 2.11 (1.44) & 2.56 (1.45) \\
distance to station origin & 1188.56 (1909.57) & 1485.12 (2496.11) & 1125.32 (2205.6) \\
distance to station destination & 1200.12 (1919.74) & 1488.22 (2497.3) & 1117.58 (2226.92) \\
origin hour & 14.18 (4.79) & 13.98 (4.43) & 14.4 (4.51) \\
origin day & 2.8 (1.89) & 2.81 (1.9) & 2.83 (1.88) \\
destination hour & 14.07 (4.93) & 13.91 (4.53) & 14.06 (4.66) \\
destination day & 2.82 (1.87) & 2.82 (1.88) & 2.85 (1.86) \\
\bottomrule
\end{tabular}
}
    \caption{Comparison of trip features of MOBIS car sharing users between different phases of the pandemic.}
    \label{tab:pre_post_covid}
\end{table}

\end{document}